%% file: BAM251_draft.tex
\documentclass[twocolumn,showpacs,superscriptaddress]{revtex4-1}
\usepackage{mymacros}
\usepackage{graphicx}
\usepackage{units}
\usepackage{multirow}
\usepackage{bigstrut}
\usepackage{overpic}
\usepackage{array}
\usepackage{amsfonts}
\usepackage{amsmath}
\usepackage{amssymb}
\usepackage{amsfonts}
\usepackage{verbatim}
\usepackage{footmisc}
\usepackage{color}
\usepackage{bm}
\usepackage{lipsum}
\usepackage{epstopdf}
\usepackage{warpcol}
\usepackage{soul}
\usepackage{hyperref}
\hypersetup{backref,
pdfpagemode=FullScreen,
colorlinks=true}

\newcommand{\modethirty}{\Lambda_c^+\to\Lambda\pi^+}
\newcommand{\modesixty}{\Lambda_c^+\to\Sigma^0\pi^+}

\newcommand{\mbc}{M_{\rm{BC}}}

\newcommand{\alphazero}{\alpha^{+}_{p\ks}}
\newcommand{\alphathirty}{\alpha^{+}_{\Lambda\pi^+}}
\newcommand{\alphasixty}{\alpha^{+}_{\Sigma^0\pi^+}}
\newcommand{\alphasixtytwo}{\alpha^{+}_{\Sigma^+\pi^0}}
\newcommand{\ks}{K_S^0}

\newcommand{\ldc}{\Lambda_c^+}
\newcommand{\ldcb}{\bar\Lambda_c^-}

\RequirePackage{lineno}
\begin{document}

%\setpagewiselinenumbers
%\modulolinenumbers[2]
%\linenumbers

\title{\large \bf \boldmath Measurements of Weak Decay Asymmetries of $\Lambda_c^+\to p\ks$, $\Lambda\pi^+$, $\Sigma^+\pi^0$, and $\Sigma^0\pi^+$}

\date{\small  \today}

\input{authors.tex}

\vspace{4cm}
\begin{abstract}

  Using $\ee\to\ldc\bar\Lambda_c^-$ production from a 567 pb$^{-1}$
  data sample collected by BESIII at 4.6 GeV, a full angular analysis
  is carried out simultaneously on the four decay modes of $\ldc\to
  pK_S^0$, $\Lambda \pi^+$, $\Sigma^+\pi^0$, and $\Sigma^0\pi^+$.
  For the first time,  the $\ldc$ transverse polarization is studied in
  unpolarized $\ee$ collisions, where a non-zero effect is observed with a statistical significance of
  2.1$\sigma$.
  The decay asymmetry parameters of the $\ldc$ weak
  hadronic decays into $p\ks$, $\Lambda\pi^+$, $\Sigma^+\pi^0$ and
  $\Sigma^0\pi^+$ are measured to be $
  0.18\pm0.43(\rm{stat})\pm0.14(\rm{syst})$,
  $-0.80\pm0.11(\rm{stat})\pm0.02(\rm{syst})$,
  $-0.57\pm0.10(\rm{stat})\pm0.07(\rm{syst})$, and
  $-0.73\pm0.17(\rm{stat})\pm0.07(\rm{syst})$, respectively.  In
  comparison with previous results, the measurements for the
  $\Lambda\pi^+$ and $\Sigma^+\pi^0$ modes are consistent but with
  improved precision, while the parameters for the $p\ks$ and
  $\Sigma^0\pi^+$ modes are measured for the first time.

%%%%%%%%%%%%%%%%%%%%%%%%%%%%%%%%%%%
\end{abstract}

\maketitle

%%%%%%%%%%%%%%%%%%%%%%%%%%%%%%%%%%%%%%%%%%%%%%%%%%%%%%%%%%%%%%%%
%%%%%     Introduction       Part                  %%%%%%%%%%%%%
%%%%%%%%%%%%%%%%%%%%%%%%%%%%%%%%%%%%%%%%%%%%%%%%%%%%%%%%%%%%%%%%

The study of the lightest charmed baryon $\ldc$ is important for the
understanding of the whole charmed baryon sector.  In recent years,
there has been significant progress in studying the $\ldc$, both
experimentally and theoretically~\cite{Amhis:2016xyh,pdg}.  This
provides crucial information in detailed explorations of the singly
charmed baryons ($\Sigma_c$, $\Xi_c$ and
$\Omega_c$)~\cite{Lu:2016ogy,Geng:2017mxn}, and further searches or
discoveries of the doubly charmed baryons ($\Xi_{cc}$ and
$\Omega_{cc}$)~\cite{Yu:2017zst,Aaij:2017ueg}. Moreover, as the
charmed baryon is the favored weak decay final state of $b$-baryons and its properties are inputs to study $b$-baryons,
improved knowledge in the charm sector can contribute substantially to
understanding the properties of $b$-baryons.

Some QCD-inspired charmed baryon models that have been
developed~\cite{Cheng:2015iom} are the flavor symmetry model
\cite{Savage:1989qr}, factorization model~\cite{Bjorken:1988ya}, pole
model~\cite{Cheng:1993gf}, and current algebra framework
~\cite{Xu:1992vc}.  As shown in Refs.~\cite{Cheng:2015iom,pdg},
many of these models calculate $\ldc$ decay rates in good agreement
with experimental results. But the decay asymmetries predicted by
these models for $\ldc$ two-body hadronic weak decays do not agree very
well.

The decay asymmetry parameter, $\alpha^+_{BP}$, in a weak decay
$\ldc\to BP$ ($B$ denotes a $J^P =\frac{1}{2}^+$ baryon and $P$ denotes a
$J^P=0^-$ pseudoscalar meson) is defined as
$\alpha^+_{BP}\equiv\frac{2{\rm Re}(s\cdot p)}{|s|^2+|p|^2}$, where
$s$ and $p$ stand for the parity-violating $s$-wave and
parity-conserving $p$-wave amplitudes in the decay,
respectively. Model calculations of $\alpha^+_{BP}$ in $\ldc\to p\ks$,
$\Lambda\pi^+$, $\Sigma^+\pi^0$, and $\Sigma^0\pi^+$ are listed in
Table~\ref{tab:theory}, which shows large variations among the
different models. As predictions of $\alpha^+_{BP}$ rely on the
relative phase between the two amplitudes, the experimental
measurements of the decay asymmetry parameters serve as very sensitive
probes to test different theoretical models. %~\cite{Asner:2008nq}.

Experimentally, only $\alphathirty$
and $\alphasixtytwo$ have been measured previously
\cite{Link:2005ft,Bishai:1995gp,Albrecht:1991vs,Avery:1990ya,Bishai:1995gp}.
The measured value for $\alphasixtytwo$ is $-0.45\pm0.32$, in
contradiction with the predicted values in many theoretical models
\cite{Cheng:1993gf, Xu:1992vc, Korner:1992wi, Cheng:1991sn,
  Ivanov:1997ra, Zenczykowski:1993jm}.  Therefore, it is important to
carry out independent measurements of $\alphasixtytwo$ to confirm the
sign of $\alphasixtytwo$ and test these models.  Moreover,
$\alphasixtytwo$ and $\alphasixty$ should have the same value
according to hyperon isospin symmetry~\cite{Sharma:1998rd}, and any
deviation from this expectation provides critical information on final
state interactions in $\ldc$ hadronic decays.  All the models predict
$\alphathirty$ consistent with the measured values, and it is
necessary to further improve the experimental precision to
discriminate between them.

In previous experiments, $\ldc$ was assumed to be unpolarized, and the
decay asymmetry parameter $\alpha^+_{BP}$ was obtained by analyzing
the longitudinal polarization from the weak two-body decay of the
produced baryon $B$, such as $\Lambda\to p \pi^-$ and $\Sigma^+\to p
\pi^0$ for $\alphathirty$ and $\alphasixtytwo$, respectively. However,
the hypothesis of unpolarized $\ldc$ may not be valid.  There have
been observations of transverse $\Lambda$ polarization in inclusive
$\Lambda$ production in $\ee$ collisions at 10.58
GeV~\cite{Guan:2018ckx} and in $\ee\to \Lambda \bar{\Lambda}$ at $J/\psi$ mass position~\cite{Lamtrans},
and it has been postulated that the produced
$\ldc$ could be polarized~\cite{Faldt:2017yqt}. Further, as the
polarization of the proton in the decay $\ldc\to p \ks$ is not
accessible with the above method, a non-zero transverse polarization
of the $\ldc$ provides an alternative way to measure
$\alphazero$~\cite{supple}.

In this Letter, we investigate for the first time the transverse
polarization of the $\ldc$ baryon in unpolarized $\ee$ annihilations.
We present for the first time measurements of the decay asymmetry
parameters in $\ldc$ decays into $p\ks$, $\Lambda\pi^+$,
$\Sigma^+\pi^0$, and $\Sigma^0\pi^+$%~\cite{ccmode}
 based on a
multi-dimensional angular analysis of the cascade-decay final states,
which greatly improves the resulting precision. Data sample used in this analysis corresponds to
an integrated luminosity of $567\,\rm{pb}^{-1}$ collected with the
BESIII detector at BEPCII at center-of-mass (CM) energy of 4.6 GeV.

Since the close proximity of the CM energy to the
$\Lambda_c^+\bar\Lambda_c^-$ mass threshold does not allow an
additional hadron to be produced, $\Lambda_c^+\bar\Lambda_c^-$ are
always generated in pairs, which provides a clean environment to study
their decays. When one $\ldc$ is detected, another
$\bar\Lambda_c^-$ partner is inferred. Hence, to increase signal
yields, we adopt a partial reconstruction method, in which only one
$\Lambda_c^+$ is reconstructed out of all the final-state particles in
an event. The charge conjugation modes are always implied in the context,
unless otherwise stated explicitly.

Details of the BESIII apparatus, the software framework and the Monte
Carlo (MC) simulation sample have been given in Ref.~\cite{lipr}.  The
$\Lambda_c^+$ signal candidates are reconstructed through the decays
into $p\ks$, $\Lambda\pi^+$, $\Sigma^+\pi^0$ and
$\Sigma^0\pi^+$. Here, the intermediate particles $\ks$, $\Lambda$,
$\Sigma^+$, $\Sigma^0$ and $\pi^0$ are reconstructed via the decays
$\ks\to\pi^+\pi^-$, $\Lambda\to p\pi^-$, $\Sigma^+\to p\pi^0$,
$\Sigma^0\to\gamma\Lambda$, and $\pi^0\to\gamma\gamma$. The event
selection criteria follow those described in Ref.~\cite{lipr}, unless
otherwise stated explicitly. To suppress the $\Lambda_c^+\to p \ks$,
$\ks\to \pi^0\pi^0$ events in the $\Sigma^+\pi^0$ candidate samples,
the invariant mass of the $\pi^0\pi^0$ system is required to be
outside the range $[400, 550]\mevcc$.

\begin{figure}[tbp!]
\centering
\includegraphics[width=0.48\linewidth]{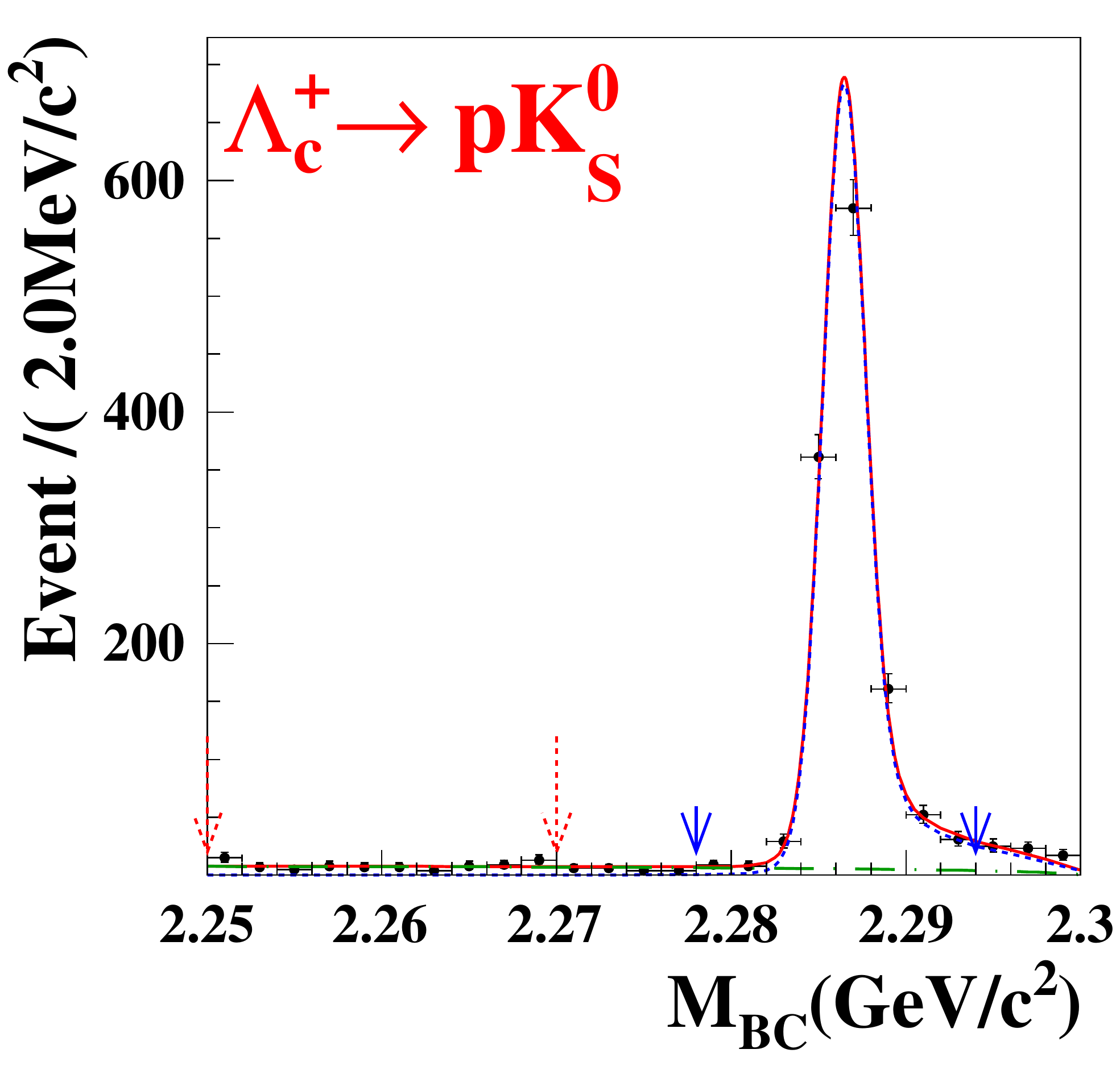}
 \put(-90,80){ (a)}
\includegraphics[width=0.48\linewidth]{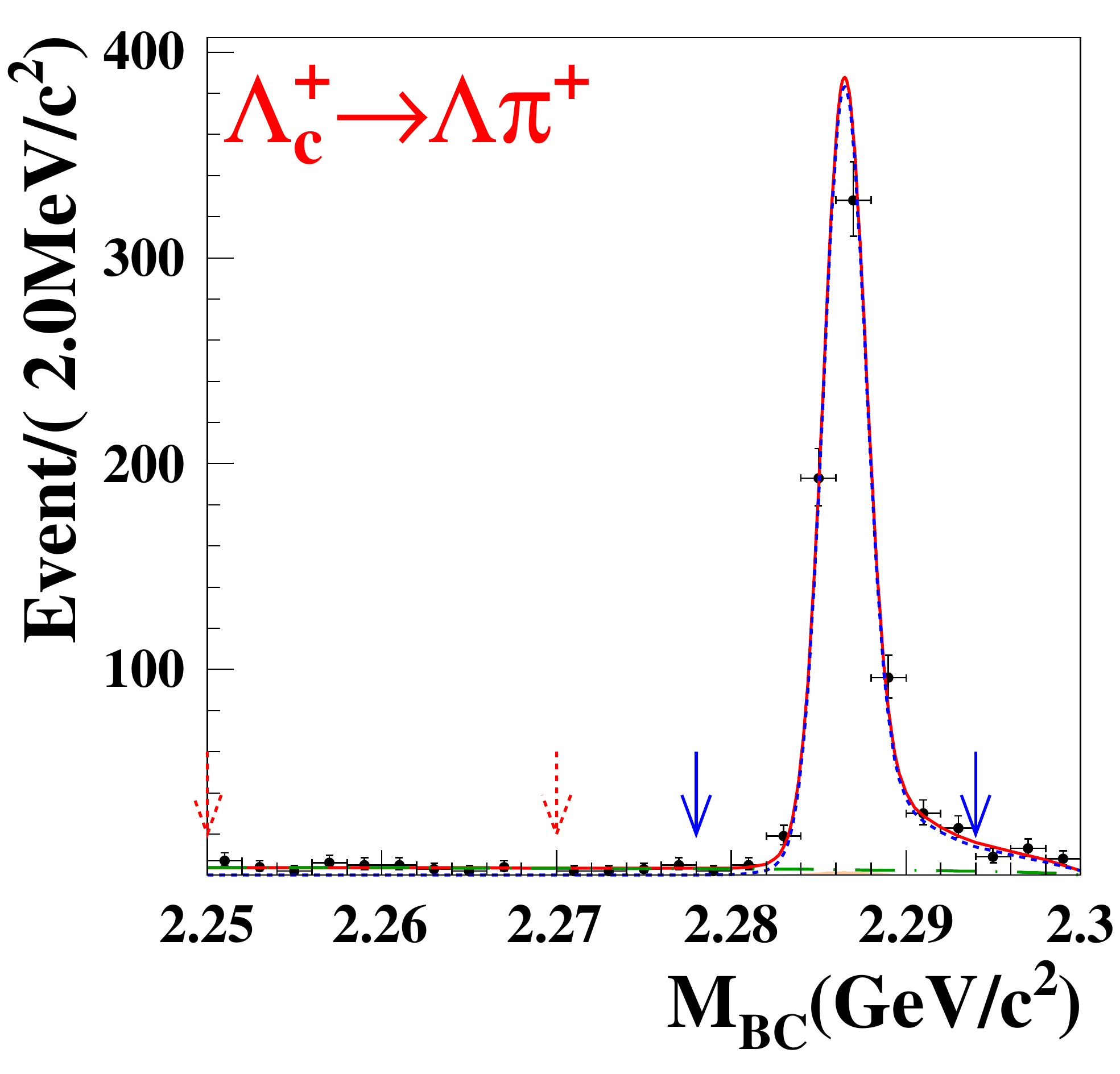}
\put(-90,80){ (b)}\\
\includegraphics[width=0.48\linewidth]{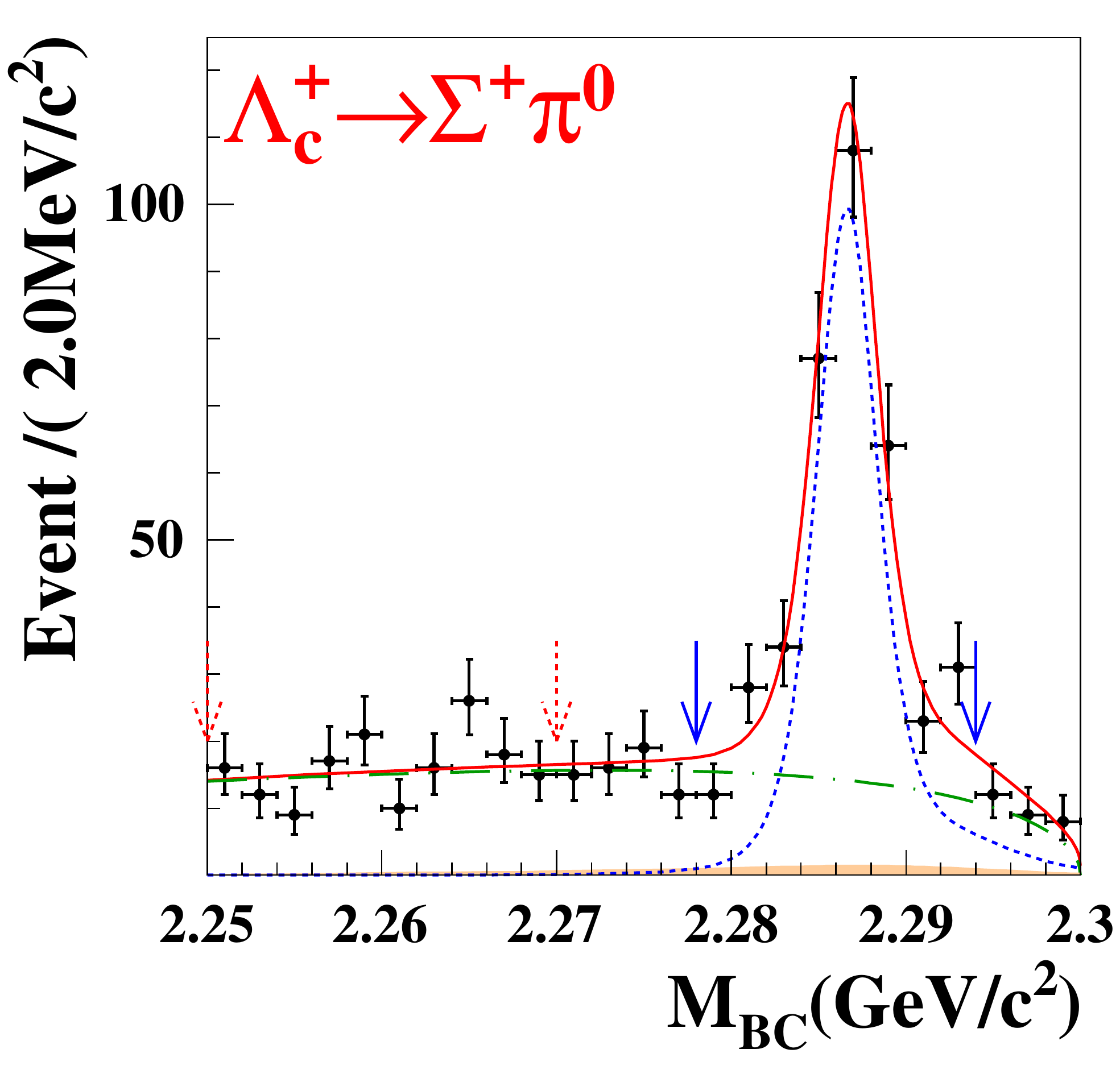}
\put(-90,80){ (c)}
\includegraphics[width=0.48\linewidth]{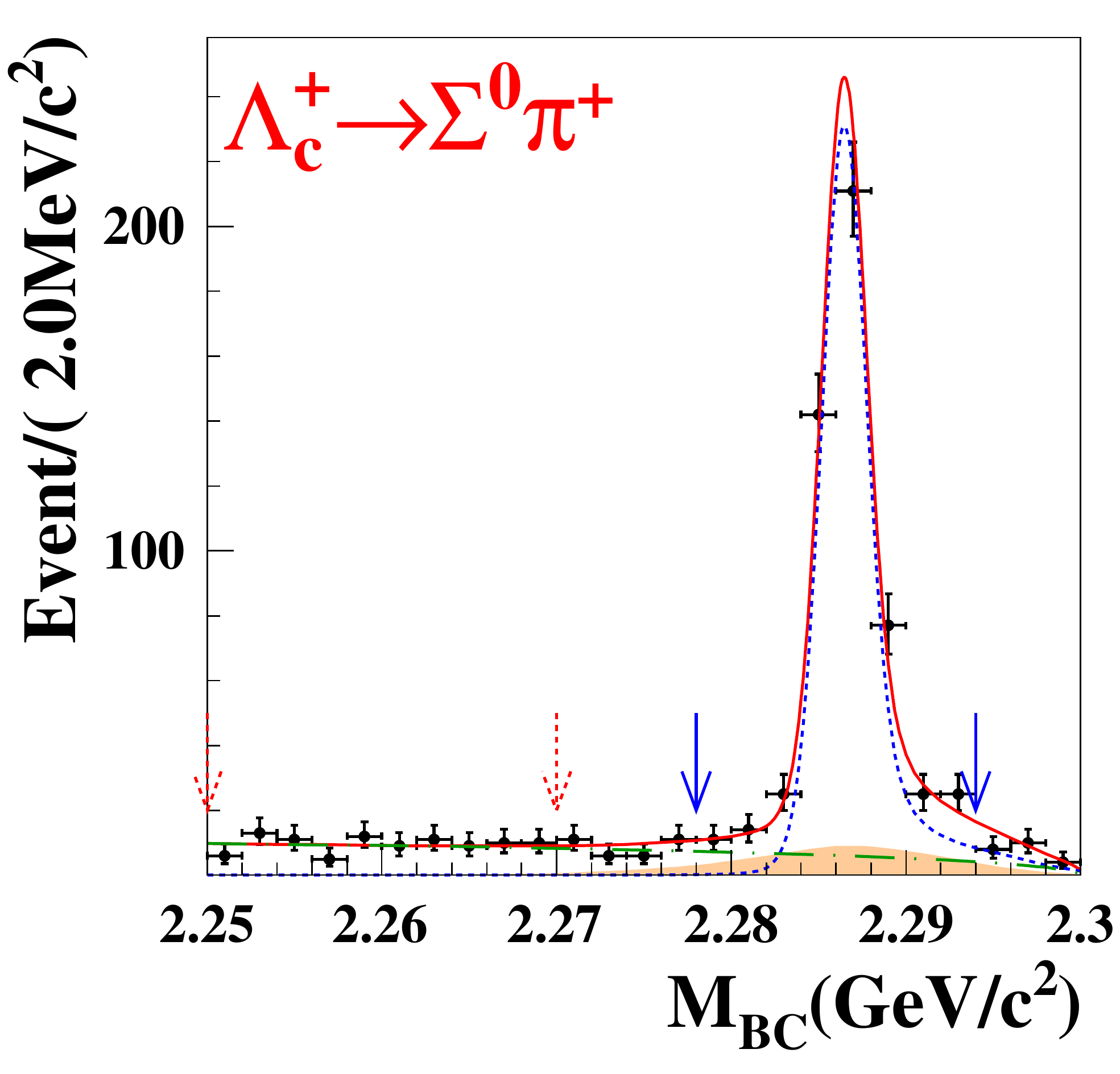}
\put(-90,80){ (d)}\\
\caption{ (color online) Fits to the $\mbc$ spectra of the signal candidates of (a) $\ldc\to p\ks$, (b) $\ldc\to\Lambda \pi^+$, (c) $\ldc\to\Sigma^+\pi^0$, and (d) $\ldc\to\Sigma^0\pi^+$.
Points with error bars correspond to data, solid lines are the fitting curves,
dashed lines describe the signal events distribution,
dash-dotted lines show the Type-II backgrounds and shadowed areas correspond to Type-I backgrounds.
Dashed and solid arrows show the sideband and signal regions, respectively.}
\label{fig:mbc}
\end{figure}

For each signal decay mode, the yields are obtained from a fit to the
beam-constrained mass ($\mbc$) distribution, $\mbc\equiv\sqrt{E^2_{\rm{beam}}-p_{\Lambda_c^+}^2}$, where $E_{\rm beam}$ is the average
beam energy and $p_{\Lambda_c^+}$ is the measured $\Lambda_c^+$
momentum in the CM system of the $e^+e^-$ collisions.  If more than
one candidate is reconstructed in the event, the one with the
smallest energy difference ($|\Delta E|$) is kept, where $\Delta E \equiv
E_{\Lambda_c^+}-E_{\rm{beam}}$, and $E_{\Lambda_c^+}$ is the measured
total energy of the $\Lambda_c^+$ candidate.

Figure \ref{fig:mbc} shows the $\mbc$ distributions for the signal
candidates, where the $\ldc$ signal peak is evident at the nominal
$\Lambda_c^+$ mass. The backgrounds can be classified into two
types. The Type-I backgrounds are from the true $\Lambda_c^+$ signal
decays, where at least one of the final state particle
candidates is wrongly assigned in reconstruction. The Type-II
backgrounds correspond to combinatorial backgrounds mostly from
$\ee\to q\bar{q}~(q=u,d,s)$ processes.  To evaluate the Type-I and
Type-II background level, unbinned maximum likelihood fits (shown in
Fig.~\ref{fig:mbc}) are applied to the $\mbc$ spectra.  The signal and
Type-I background shapes, as well as the ratio of their yields, are
derived from the signal MC simulation samples. These two shapes are
convolved with a common Gaussian function, whose width is left free
and represents the difference in resolution between data and MC
simulations. The Type-II background shape is modeled by an ARGUS
function~\cite{Albrecht:1990am}. The $\ldc$ signal and sideband regions
are chosen as $[2.278, 2.294]\,\rm{GeV}/\emph{c}^2$ and $[2.250,
2.270]\,\rm{GeV}/\emph{c}^2$, respectively.

The decay asymmetry parameters are determined by analyzing the
multi-dimensional angular distributions, where the full cascade decay
chains are considered. The full angular dependence formulae (4), (6),
and (10) in Ref.~\cite{supple}, constructed under the helicity basis,
are used in the fit. To illustrate the helicity system defined in this
analysis, we take as an example the two-level cascade decay process
$\Lambda_c^+\to\Lambda\pi^+,\Lambda\to p\pi^-$ following the level-0
process $e^+e^-\to\gamma^*\to\Lambda_c^+\bar\Lambda_c^-$. An analogous
formalism is applied to the other $\ldc\to BP$ decays.

%%%%%%%%%%%%%%%%%%
\begin{figure}[thp!]
\centering
\includegraphics[width=0.90\linewidth]{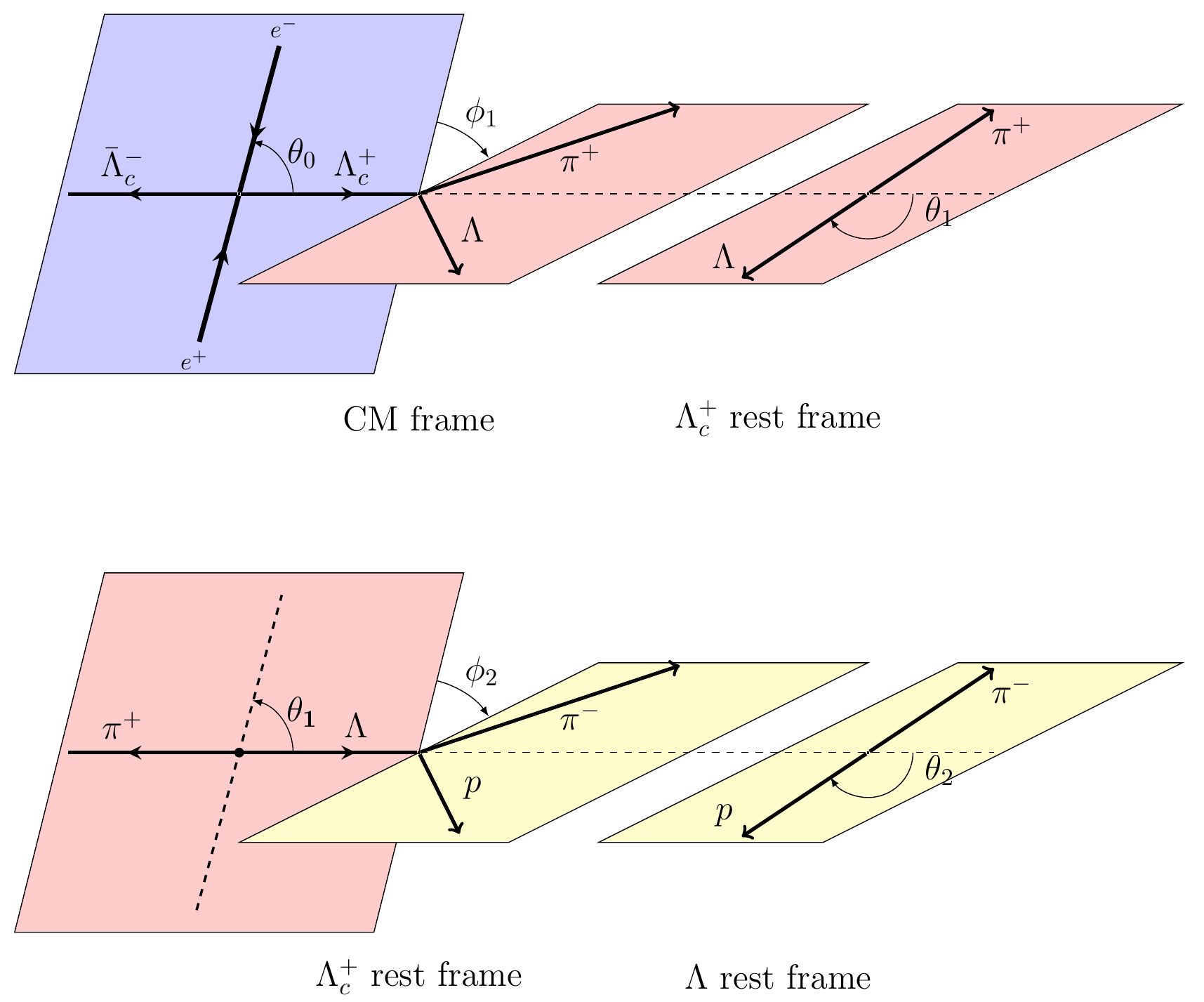}
\caption{ (color online) Definition of the helicity frame for $\ee\to\ldc\bar\Lambda_c^-$, $\ldc\to\Lambda\pi^+$, $\Lambda\to p\pi^-$ }.
\label{fig:huatu}
\end{figure}
%%%%%%%%%%%%%%%%%%%%%

Figure \ref{fig:huatu} illustrates the definitions of the full system
of helicity angles for the $\ldc\to\Lambda\pi^+$ mode.  In the
helicity frame of $e^+e^-\to\Lambda_c^+\bar\Lambda_c^-$, $\theta_0$ is
the polar angle of the $\Lambda_c^+$ with respect to the $e^+$ beam
axis in the $\ee$ CM system.  For the helicity angles of the
$\ldc\to\Lambda\pi^+$ decay, $\phi_1$ is the angle between the
$e^+\ldc$ and $\Lambda\pi^+$ planes, and $\theta_1$ is the polar angle
of the $\Lambda$ momentum in the rest frame of the $\ldc$ with respect
to the $\ldc$ momentum in the CM frame.  The angle subscript
represents the level numbering of the cascade signal decays.  For the
helicity angles describing the $\Lambda\to p \pi^+$ decay, $\phi_2$ is
the angle between the $\Lambda\pi^+$ plane and $p\pi^-$ plane and
$\theta_2$ is the polar angle of the proton momentum with respect to
opposite direction of $\pi^+$ momentum in the rest frame of
$\Lambda$. For the three-level cascade decays
$\ldc\to\Sigma^0\pi^+,~\Sigma^0\to\Lambda\gamma,~\Lambda\to p\pi^-$
process, $\phi_3$ is the angle between the $\Lambda\gamma$ and
$p\pi^-$ planes, while $\theta_3$ is the polar angle of the proton
with respect to the opposite direction of the photon momentum (from
$\Sigma^0\to \Lambda \gamma$) in the rest frame of $\Lambda$.

In Ref.~\cite{supple}, we define $\Delta_0$ as the phase angle difference between two individual helicity amplitudes, $H_{\lambda_1,\lambda_2}$, for the $\ldc$ production process $\gamma^*\to\ldc(\lambda_1)\bar\Lambda_c^-(\lambda_2)$ with total helicities $|\lambda_1-\lambda_2| = 0$ and $1$, respectively.
In the case where one-photon exchange dominates the production process, $\Delta_0$ is also the phase between the electric and magnetic form factors of the $\ldc$~\cite{Ablikim:2017lct,Faldt:2017yqt}. The transverse polarization observable of the produced $\ldc$ can be defined as
\begin{equation}
\label{eq_pt}
\mathcal{P}_T(\cos\theta_0)\equiv\sqrt{1-\alpha_0^2}\cos\theta_0\sin\theta_0\sin\Delta_0,
\end{equation}
whose magnitude varies as a function of $\cos\theta_0$.
Similarly, two parameters, $\alpha^+_{BP}$ and $\Delta_1^{BP}$, describe the level-1 decays $\ldc \to\Lambda \pi^+,~\Sigma^+\pi^0$, and $\Sigma^0\pi^+$, where $\Delta_1^{BP}$ is the phase angle difference between the two helicity amplitudes in the $BP$ mode.
The Lee-Yang parameters~\cite{Lee:1956qn,supple} can be obtained with the relations
\begin{equation}
\begin{array}{ll}
\beta_{BP} & = \sqrt{1-(\alpha^{+}_{BP})^2}\sin\Delta_1^{BP}, \\
\gamma_{BP}& = \sqrt{1-(\alpha^{+}_{BP}})^2\cos\Delta_1^{BP}. \\
\end{array}
\end{equation}

In the angular analysis, the free parameters describing the angular distributions for the four data sets
are determined from a simultaneous unbinned maximum likelihood fit, as $\alpha_0$ and $\Delta_0$ are common.
The likelihood function is constructed from the probability density function (PDF) jointly by
\begin{equation}
  \mathcal{L}_{\rm data} = \prod_{i=1}^{N_{\rm data}}f_S(\vec\xi).
\end{equation}
Here, $f_S(\vec\xi)$ is the PDF of the signal process, $N_{\rm data}$ is the number of the events in data and $i$ is event index. Signal PDF $f_S(\vec\xi)$ is formulated as
\begin{equation}
  f_S(\vec\xi)=\frac{\epsilon(\vec{\xi})|M(\vec\xi; \vec\eta)|^2}{\int\epsilon(\vec\xi)|M(\vec\xi; \vec\eta)|^2{\rm d}\vec\xi}~,
\end{equation}
where the variable $\vec\xi$ denotes the kinematic angular observables, and $\vec\eta$ denotes the free parameters to be determined. $M(\vec\xi)$ is the total decay amplitude~\cite{supple} and $\epsilon(\vec\xi)$ is the detection efficiency parameterized in terms of the kinematic variables $\vec\xi$.
The background contribution to the joint likelihood is subtracted according to the calculated likelihoods for the Type-I
background based on inclusive MC simulations and for the Type-II background according to the $M_{\rm BC}$ sideband.
The MC-integration technique is adopted to compute the
normalization factor as follows
\begin{equation}
  \int\epsilon(\vec\xi)|M(\vec\xi; \vec\eta)|^2{\rm d}\vec\xi=\frac{1}{N_{\rm gen}}\sum^{N_{\rm MC}}_{k_{\rm MC}}|M(\vec\xi_k; \vec\eta)|^2,
\end{equation}
where $N_{\rm gen}$ is the total number of MC-simulated signal events. $N_{\rm MC}$ is the number of the MC signal events survived from the full selection criteria and $k_{\rm MC}$ is its event index.

Minimization of the negative logarithmic likelihood with background subtraction over all the
four signal processes is carried out using the {\sc{MINUIT}} package
\cite{James:1975dr}.
Here, $\alpha_0$ is
fixed to the known value $-0.20$ \cite{Ablikim:2017lct}.  For the
charge-conjugation $\ldcb$ decays, under the assumption of $CP$
conservation, $\bar\Delta_0=\Delta_0$,
$\alpha^{+}_{BP}=-\alpha^{-}_{\bar B\bar P}$, and $\bar
\Delta_1^{\bar{B}\bar{P}}=-\Delta_1^{BP}$. The decay asymmetry
parameter $\alpha_\Lambda$ for $\Lambda\to p \pi^-$ is taken from the
recent BESIII measurement~\cite{Lamtrans} and $\alpha_{\Sigma^+}$ for
$\Sigma^+\to p \pi^0$ from the Particle Data Group (PDG)~\cite{pdg}.
From the fit, we obtain
$\sin\Delta_0=-0.28\pm0.13({\rm stat.})$ which differs from zero with
a statistical significance of 2.1$\sigma$ according to a likelihood
ratio test.  This indicates that transverse polarization
$\mathcal{P}_T$ of the $\ldc$ is non-zero when $\sin(2\theta_0)\ne 0$.
The numerical fit results are given in Table \ref{tab:theory},
together with the calculated $\gamma_{BP}$ and $\beta_{BP}$.

\begin{table*}[tp!]%[tbp!]
  \centering
  \caption{Comparisons between different theoretical calculations and experimental measurements. }
  \label{tab:theory}
  \begin{tabular}{l|l|ll|ll|ll|ll}
  \hline
  \hline
   $\ldc\to$  &~& \multicolumn{2}{c}{ $p\ks$}    & \multicolumn{2}{c}{$\Lambda\pi^+$} & \multicolumn{2}{c}{$\Sigma^+\pi^0$} & \multicolumn{2}{c}{$\Sigma^0\pi^+$}\\
  \hline
  \multirow{7}*{$\alpha_{BP}^{\ldc}$}& \multirow{5}*{Predicted} &$-1.0$ \cite{Korner:1992wi}, & $~~\,0.51$ \cite{Xu:1992vc} & $-0.70$ \cite{Korner:1992wi}, &$-0.67$ \cite{Xu:1992vc}        & $~~\,0.71$ \cite{Korner:1992wi},& $~~\,0.92$ \cite{Xu:1992vc}            & $~~\,0.70$ \cite{Korner:1992wi},& $~~\,0.92$ \cite{Xu:1992vc}   \\

  ~&~& $-0.49$ \cite{Cheng:1993gf},& $-0.90$ \cite{Cheng:1993gf}& $-0.95$ \cite{Cheng:1993gf},& $-0.99$ \cite{Cheng:1993gf}   & $~~\,0.79$ \cite{Cheng:1993gf}  &     $-0.49$ \cite{Cheng:1993gf}           & $~~\,0.78$ \cite{Cheng:1993gf},&  $-0.49$ \cite{Cheng:1993gf}     \\

  ~&&$-0.49$ \cite{Cheng:1991sn}, & $-0.97$ \cite{Ivanov:1997ra}& $-0.96$ \cite{Cheng:1991sn},& $-0.95$ \cite{Ivanov:1997ra}& $~~\,0.83$ \cite{Cheng:1991sn}, &$~~\,0.43$ \cite{Ivanov:1997ra}            & $~~\,0.83$ \cite{Cheng:1991sn},& $~~\,0.43$ \cite{Ivanov:1997ra}   \\

  ~&~&$-0.66$ \cite{Zenczykowski:1993jm}, & $-0.90$ \cite{Zenczykowski:1993hw}   & $-0.99$ \cite{Zenczykowski:1993jm}, &$-0.86$ \cite{Zenczykowski:1993hw}     & $~~\,0.39$ \cite{Zenczykowski:1993jm},& $-0.76$ \cite{Zenczykowski:1993hw} & $~~\,0.39$ \cite{Zenczykowski:1993jm}, &$-0.76$ \cite{Zenczykowski:1993hw}   \\

  ~&~&$-0.99$ \cite{Sharma:1998rd},  &$-0.91$ \cite{Datta:1995mn}    & $-0.99$ \cite{Sharma:1998rd},& $-0.94$ \cite{Datta:1995mn}       & $-0.31$ \cite{Sharma:1998rd},& $-0.47$ \cite{Datta:1995mn}           & $-0.31$ \cite{Sharma:1998rd},& $-0.47$ \cite{Datta:1995mn}  \\
  \cline{2-10}
  ~&PDG \cite{pdg}&  \multicolumn{2}{c}{$ $} \vline& \multicolumn{2}{c}{$-0.91\pm0.15$} \vline&  \multicolumn{2}{c}{$-0.45\pm0.32$}   \vline&         \\
  ~&This work  & \multicolumn{2}{c}{$0.18\pm0.43\pm0.14$}  \vline& \multicolumn{2}{c}{$-0.80\pm0.11\pm0.02$}  \vline& \multicolumn{2}{c}{$-0.57\pm0.10\pm0.07$} \vline& \multicolumn{2}{c}{$-0.73\pm0.17\pm0.07$}\\
  \hline

  $\Delta^{BP}_1 ({\rm rad})$                        &This work &     \multicolumn{2}{c}{$ $}\vline& \multicolumn{2}{c}{$ 3.0\pm2.4\pm1.0$}\vline & \multicolumn{2}{c}{$4.1\pm1.1\pm0.6$} \vline& \multicolumn{2}{c}{$0.8\pm1.2\pm0.2$} \\
  $\beta_{BP}$                        &This work &     \multicolumn{2}{c}{$ $}\vline& \multicolumn{2}{c}{$ 0.06^{+0.58+0.05}_{-0.47-0.06}$}\vline & \multicolumn{2}{c}{$-0.66^{+0.46+0.22}_{-0.25-0.02}$} \vline& \multicolumn{2}{c}{$0.48^{+0.35+0.07}_{-0.57-0.13}$} \\
  $\gamma_{BP}$                       &This work &     \multicolumn{2}{c}{$ $}\vline& \multicolumn{2}{c}{$ -0.60^{+0.96+0.17}_{-0.05-0.03}$}\vline & \multicolumn{2}{c}{$-0.48^{+0.45+0.21}_{-0.42-0.04}$} \vline& \multicolumn{2}{c}{$0.49^{+0.35+0.07}_{-0.56-0.12}$} \\
  \hline
  \hline
  \end{tabular}
\end{table*}

\begin{figure}[tp!]
%  \centering
  \includegraphics[width=0.49\linewidth]{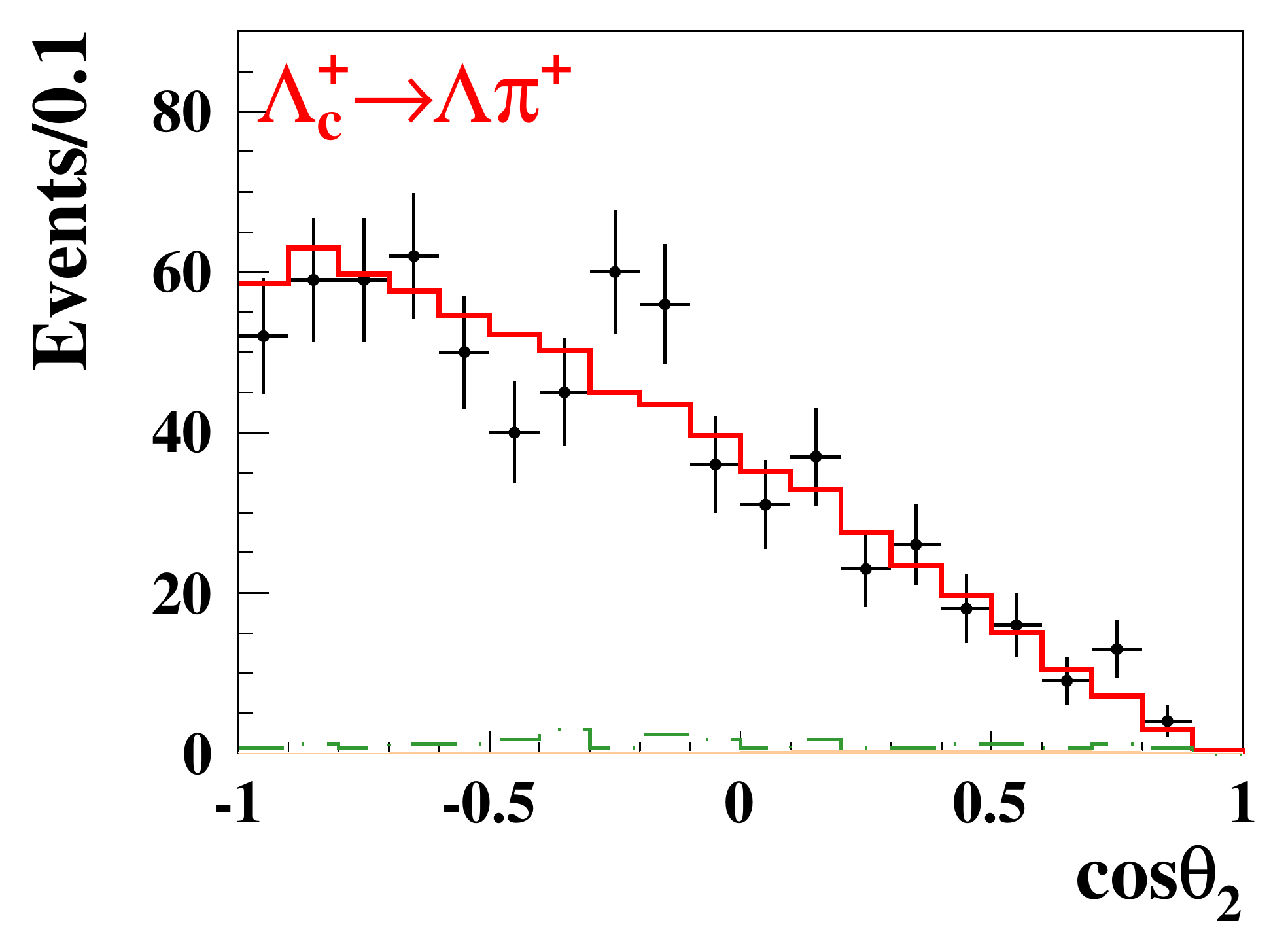}
   \put(-40,71){ (a)}
  \includegraphics[width=0.49\linewidth]{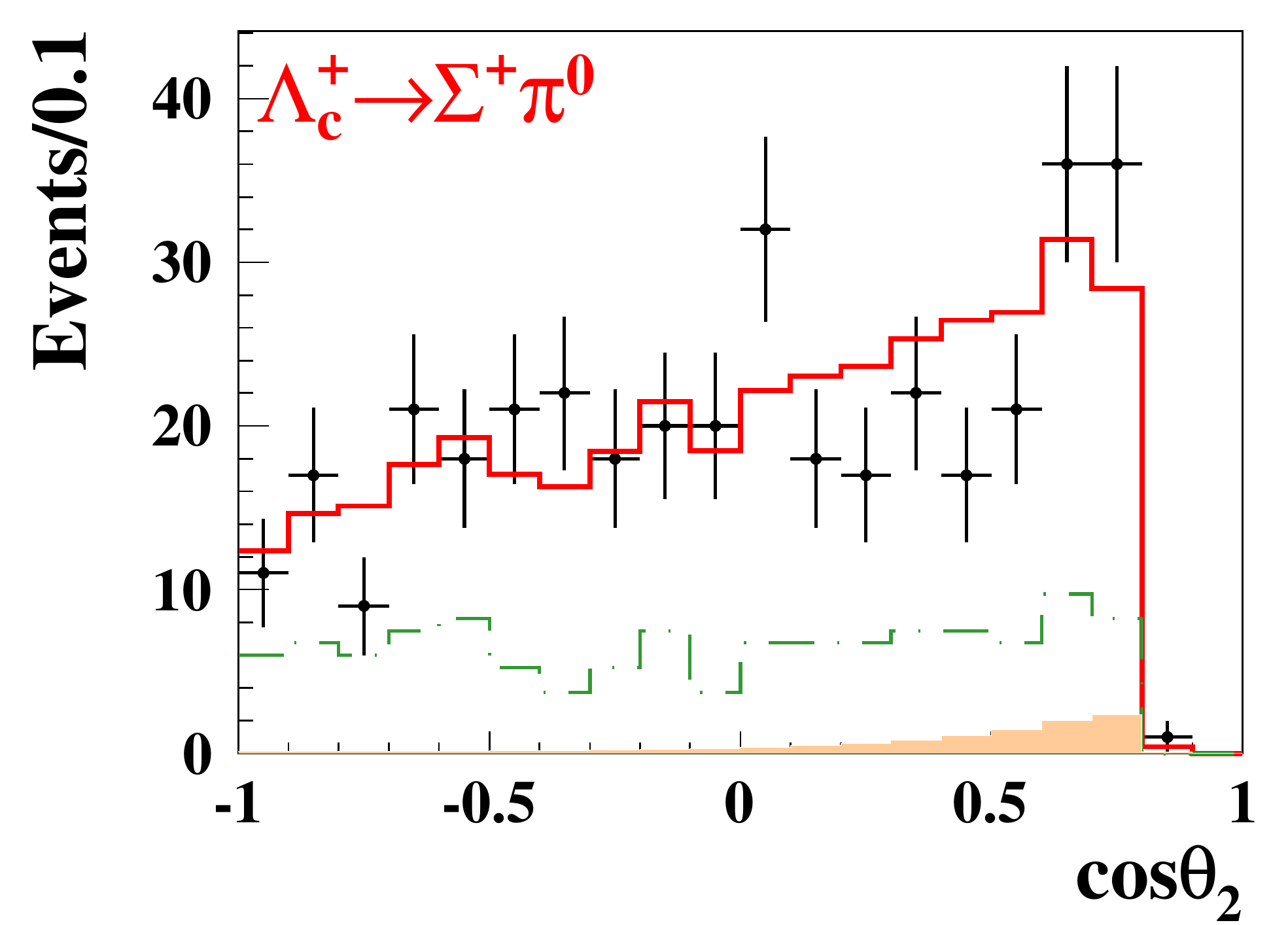}
   \put(-40,71){ (b)}\\
  \includegraphics[width=0.49\linewidth]{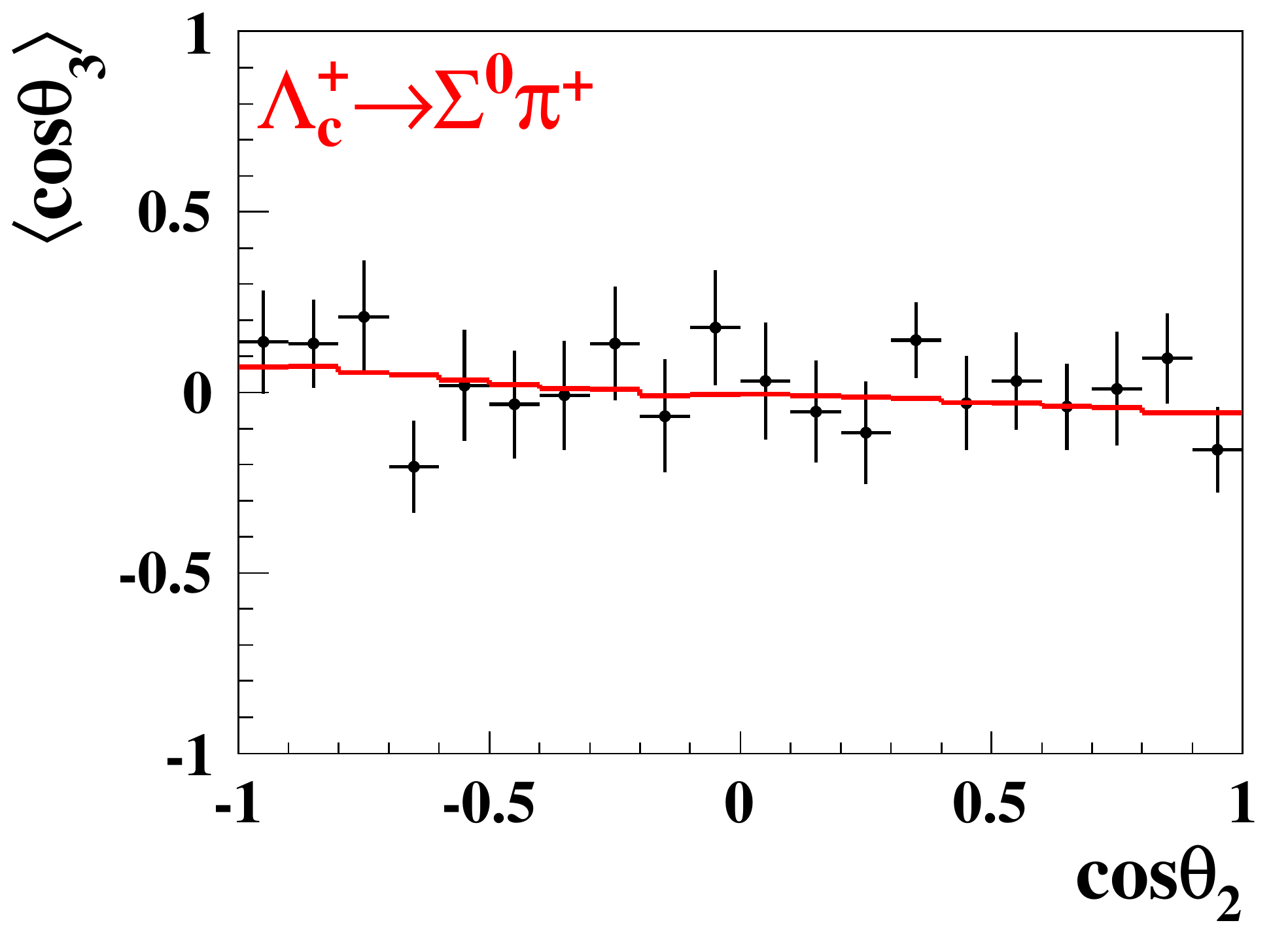}
  \put(-40,71){ (c)}
  \includegraphics[width=0.49\linewidth]{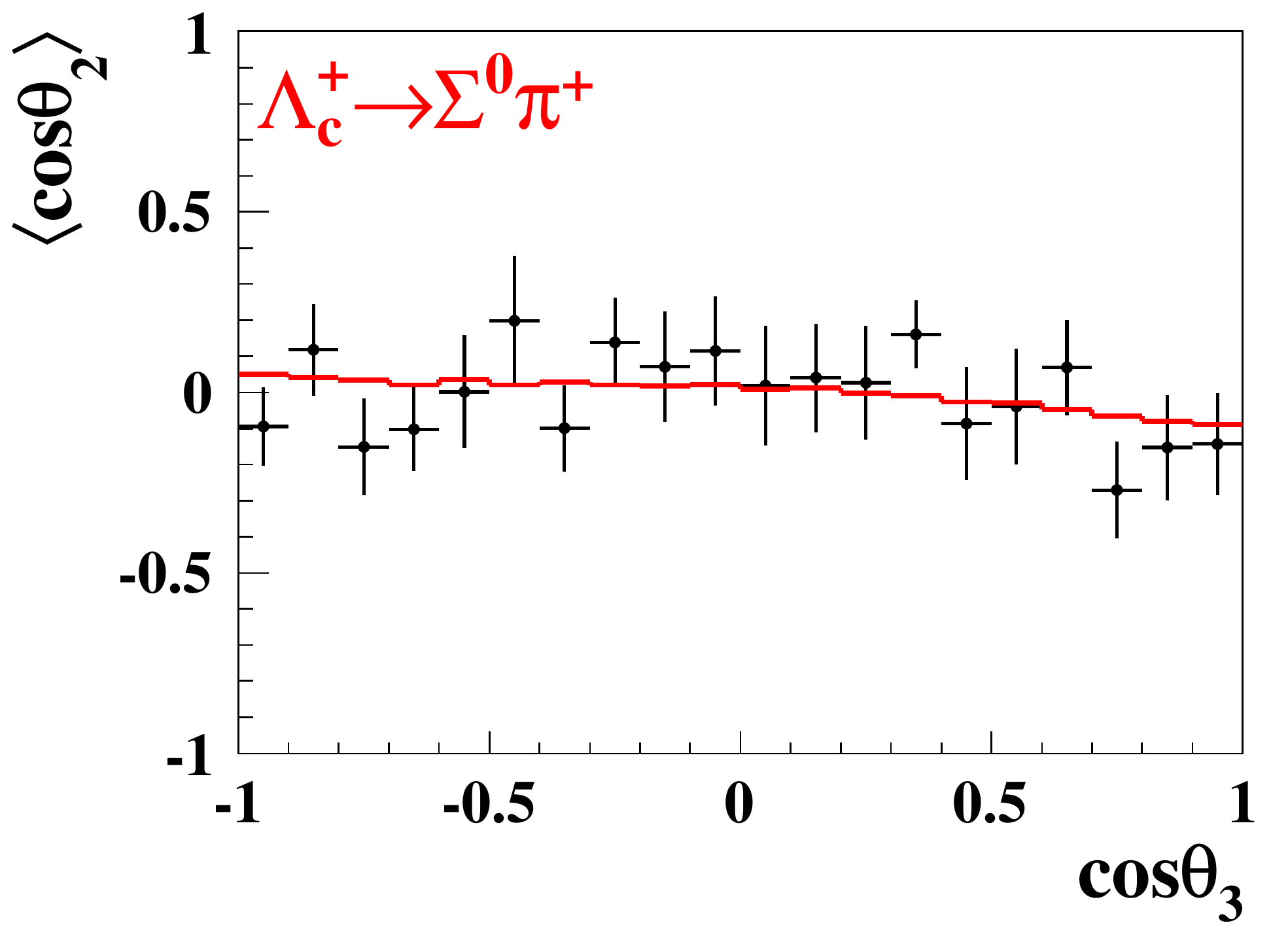}
   \put(-40,71){ (d)}\\
   \includegraphics[width=0.49\linewidth]{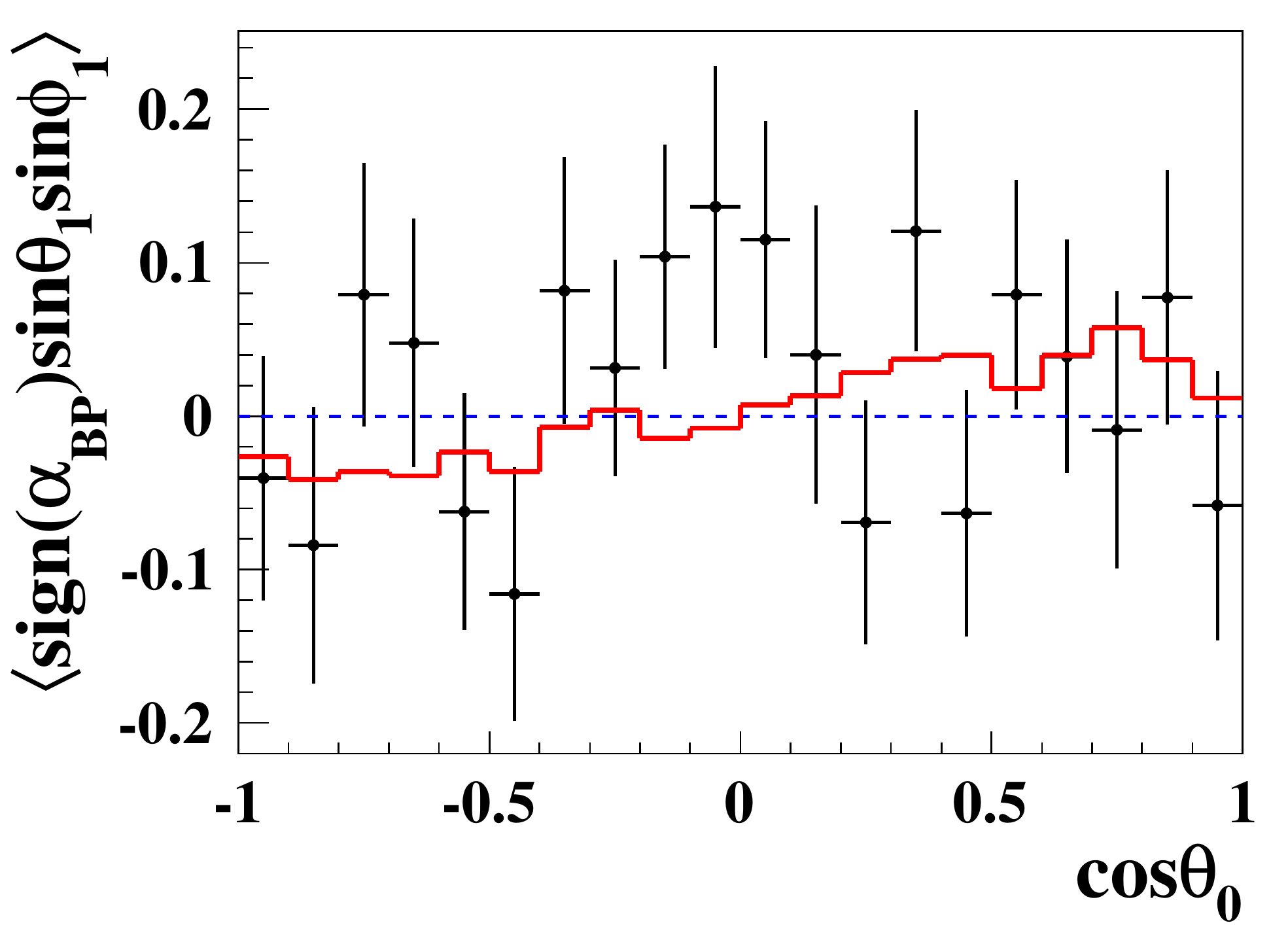}
   \put(-30,71){ (e)}\\
  \caption{(color online) $\cos\theta_2$ distributions in (a) $\Lambda\pi^+$, and (b) $\Sigma^+\pi^0$;
  (c) average value of $\cos\theta_3$ as a function of $\cos\theta_2$, and
  (d) average value of $\cos\theta_2$ as a function of $\cos\theta_3$ in $\modesixty$;
  (e)  $\<\rm{sign}(\alpha_{BP})\sin\theta_1\sin\phi_1\>$ as a function of $\cos \theta_0$ for all the four signal channels.
  Points with error bars correspond to data; (red) solid lines represent the MC-determined shapes taking into account the fit results; (green) dash-dotted lines represent the Type-II background and shaded histograms show the Type-I  background.}\label{result plot}
\end{figure}

In Fig.~\ref{result plot}, the fit results are illustrated using several projection variables. The real data are compared with the MC generated events re-weighted according to the fit.

For the $\modethirty$ and $\Sigma^+\pi^0$ decays, if all angles are
integrated over except for the angle $\theta_2$, the decay rate
becomes~\cite{wangdan}
\begin{equation}
\label{eq_the2}
  \frac{dN}{d\cos\theta_2}\propto1 + \alpha^{+}_{\Lambda\pi^+(\Sigma^+\pi^0)}\alpha_{\Lambda(\Sigma^+)}\cos\theta_2.
\end{equation}
Equation~\eqref{eq_the2} shows a characteristically  longitudinal polarization of the produced $\Lambda$($\Sigma^+$) from the $\ldc$ decays, and the asymmetry of $\cos\theta_2$ distribution reflects the product of the decay asymmetries $\alpha^{+}_{\Lambda\pi^+}\alpha_{\Lambda}$($\alpha^{+}_{\Sigma^+\pi^0}\alpha_{\Sigma^+}$)~\cite{Asner:2008nq}.
The distributions of $\cos\theta_2$ in the  $\modethirty$ and $\Sigma^+\pi^0$ modes are shown in Figs.~\ref{result plot}(a) and (b), respectively. The drop at the right side in Fig.~\ref{result plot}(b) is due to the $\ks\to\pi^0\pi^0$ veto.

For the $\modesixty$ decay, the correlations of $\cos\theta_2$ and $\cos\theta_3$ in the subsequent level-2 decay $\Sigma^0\to\gamma \Lambda$ and level-3 decay $\Lambda\to p \pi^-$, are shown in Figs. \ref{result plot}(c) and (d), respectively. The correlation of the average value of $\cos\theta_i$ satisfies the relation
\begin{equation}
\langle\cos\theta_i\rangle =-{1\over 6}\alphasixty\alpha_\Lambda\cos\theta_j ,
\end{equation}
with $(i,j)$=(2, 3) or (3, 2).

%%%%%%%%%%%%
%%%%%%%%%%%%
If the full expressions for the joint angular distributions
(Ref.~\cite{supple}) are integrated over the angles of the level 2 and
3 decay products, the remaining partial decay rate $\mathcal{W}$ is
\begin{equation}
\mathcal{W} \propto 1+\alpha_0\cos^2\theta_0 +\mathcal{P}_T
\alpha^{+}_{BP}\sin\theta_1\sin\phi_1 \label{W_eq}.
\end{equation}
Therefore, in a given $\cos\theta_0$ interval,
$$\<\sin\theta_1\sin\phi_1\>=\frac{ \int^{2 \pi}_0  \int^{1}_{-1}  \sin \theta_1 \sin \phi_1 \mathcal{W}
{\rm d}\cos \theta_1 {\rm d}\phi_1}
{ \int^{2 \pi}_0  \int^{1}_{-1} \mathcal{W}
{\rm d}\cos \theta_1 {\rm d}\phi_1 }$$
is directly proportional to
$\alpha_{BP}P_T(\cos\theta_0)/(1+\alpha_0\cos^2\theta_0)$ for the
acceptance corrected data.  In Fig.~\ref{result plot}(e), the effect
of the transverse polarization $P_T(\cos\theta_0)$ is illustrated by
plotting the average value
$\<{\rm{sign}}(\alpha_{BP})\sin\theta_1\sin\phi_1\>$ from all four
decay modes and including both particles and antiparticles. The sign
function of the measured decay asymmetry parameter, ${\rm
  sign}(\alpha_{BP})$, is used to avoid the cancellation of
contributions from the opposite charge modes.

%%%%%%%%%%%%%%%%%%%%%%%%%%%%%%%%%%%%%%%%%%%%%%%%%%%%%%
%%%%%    Systematic   uncertainty  Part  %%%%%%%%%%%%%
%%%%%%%%%%%%%%%%%%%%%%%%%%%%%%%%%%%%%%%%%%%%%%%%%%%%%%

The systematic uncertainties arise mainly from the reconstruction of final state tracks,
$K_S^0\to\pi^0\pi^0$ veto, $\Delta E$ requirement, signal $\mbc$  selections and background subtraction.
The contributions are summarized in Table \ref{systematic uncertainty}. The uncertainty of the input $\alpha_0$ is found to be negligible, after considering the experimental uncertainty \cite{Ablikim:2017lct}.
Systematic uncertainties from different sources are combined in quadrature to obtain the total systematic uncertainties.

%%%%%%%%%%%
\begin{table}[tp!]
\centering
\caption{Summary of the systematic uncertainties. $A$, $B$, $C$ and $D$ stand for the modes of $p\ks$, $\Lambda\pi^+$, $\Sigma^+\pi^0$, and $\Sigma^0\pi^+$, respectively.}
\begin{tabular}{lcccccccc}
\hline
\hline
Source             &$\alpha^+_{A}$&$\alpha^+_{B}$&$\alpha^+_{C}$&$\alpha^+_{D}$&$\sin\Delta_0$ & $\Delta_1^B$& $\Delta_1^C$ & $\Delta_1^D$\\
\hline
Reconstruction             &0.00        &    0.00     &     0.00   &     0.01   &0.00 & 0.8 & 0.0 & 0.0\\
$\pi^0\pi^0$ veto     &0.01        &    0.00     &     0.01   &     0.00   &0.00 & 0.0 & 0.2 & 0.0 \\
$\Delta E$ signal region         &0.07        &    0.01     &     0.02   &     0.05   &0.02 & 0.3 & 0.1 & 0.1\\
$\mbc$  signal region          &0.12        &    0.01     &     0.05   &     0.02   &0.02 & 0.5 & 0.4 &0.1\\
Bkg subtraction      &0.03        &    0.01     &     0.05   &     0.04   &0.02 & 0.3 & 0.3 & 0.0\\
\hline
$\textbf{Total}$      &0.14        &    0.02     &     0.07   &     0.07   &0.03 &1.0 & 0.6 &0.2\\
\hline
\end{tabular}
\label{systematic uncertainty}
\end{table}
%%%%%%%%%%%%%%%%%

To understand the reconstruction efficiencies in data and MC
simulations, a series of control samples are used for different final
states. The proton and charged pion are studied based on the channel
$J/\psi\to p\bar{p}\pi^+\pi^-$, photon on $\ee\to \gamma
\mu^+\mu^-$~\cite{Prasad:2016wxl}, $\pi^0$ on $\psi(3686)\to
\pi^0\pi^0 J/\psi$ and $\ee\to \omega\pi^0$, $\Lambda$ on
$J/\psi\to\bar pK^+\Lambda$ and
$J/\psi\to\Lambda\bar\Lambda$~\cite{Ablikim:2018jfs}, and $K_S^0$ on
$J/\psi\to K^*(892)^+K^-, K^*(892)^+\to K_S^0\pi^+$ and $J/\psi\to\phi
K_S^0K^+\pi^-$~\cite{Ablikim:2015qgt}.  The efficiency differences
between data and MC simulations are used to reweight the summed
likelihood values. The changes of the fit results after likelihood
minimization are taken as systematic uncertainties. The uncertainties
due to the $K_S^0\to\pi^0\pi^0$ veto in $\Sigma^+\pi^0$ candidate
events are evaluated by taking the maximum changes with respect to the
nominal results when varying the $\pi^0\pi^0$ veto range.  A similar
method is applied when estimating the systematic uncertainties from
the signal $\Delta E$ and $\mbc$ selection criteria.  In the
likelihood construction, the subtraction of the background
contributions are modeled with the sideband control samples and the
inclusive MC samples. The associated uncertainties are studied by
varying the sideband range and adjusting the scaling factors of the
two background components.  The altered scaling factors are obtained
by changing the background lineshapes within their 1$\sigma$
uncertainties from the fits to the $\mbc$ distribution.  The resultant
maximum changes of the fit results are taken as corresponding
systematic uncertainties.

%%%%%    summary       Part                %%%%%%%%%%%%%

To summarize, based on the 567 pb$^{-1}$ data sample collected from
$\ee$ collisions at a CM energy of 4.6 GeV, a simultaneous
full angular analysis of four decay modes of $\ldc\to pK_S^0$,
$\Lambda \pi^+$, $\Sigma^+\pi^0$, and $\Sigma^0\pi^+$ from the
$\ee\to\ldc\bar\Lambda_c^-$ production is carried out. We study the $\ldc$ transverse polarization in unpolarized $\ee$
collisions for the first time, which gives $\sin\Delta_0=-0.28\pm0.13\pm0.03$
with a statistical significance of 2.1$\sigma$.
This information will help in
understanding the production mechanism of the charmed baryons in $\ee$
annihilations.  With availability of  the transverse polarization measurement, the decay
asymmetry parameter in $\ldc\to pK_S^0$ becomes accessible
experimentally.  Moreover, this improves
the precision in determining the decay asymmetry parameters in
$\ldc\to\Lambda \pi^+$, $\Sigma^+\pi^0$, and $\Sigma^0\pi^+$, as
listed in Table~\ref{tab:theory}.

The parameters $\alphazero$ and
$\alphasixty$ are measured for the first time. The measured
$\alphathirty$ and $\alphasixtytwo$ parameters are consistent with
previous measurements, but with much improved precisions (by a factor
of 3 for $\alphasixtytwo$).  The negative sign of the $\alphasixtytwo$
parameter is confirmed and differs from the positive
predictions~\cite{Korner:1992wi, Xu:1992vc, Cheng:1993gf,
  Cheng:1991sn, Ivanov:1997ra, Zenczykowski:1993jm} by at least
8$\sigma$, which rules out those model calculations.  The measured
$\alphasixtytwo$ and $\alphasixty$ values agree well, which supports
hyperon isospin symmetry in $\ldc$ decay.  For the results on $\alphazero$,
$\alphasixtytwo$, and $\alphasixty$ listed in Table~\ref{tab:theory},
at present no model gives predictions fully consistent with all the
measurements.
These improved results in $\ldc$ decay asymmetries provide essential inputs for the
$b$-baryon decay asymmetry measurements to be performed in the future.

%%%%%%%%%%%%%%%%%%%%%%%%%%%%%%%%%%%%%%%%%%%%%%%%%%%%%%%%%%%%%%%%
%%%%%    acknowledgments       Part                %%%%%%%%%%%%%
%%%%%%%%%%%%%%%%%%%%%%%%%%%%%%%%%%%%%%%%%%%%%%%%%%%%%%%%%%%%%%%%
\vspace{1.0cm}

The BESIII collaboration thanks the staff of BEPCII and the IHEP computing center for their strong support. This work is supported in part by National Key Basic Research Program of China under Contract No. 2015CB856700; National Natural Science Foundation of China (NSFC) under Contracts Nos. 11335008, 11425524, 11625523, 11635010, 11735014; the Chinese Academy of Sciences (CAS) Large-Scale Scientific Facility Program; the CAS Center for Excellence in Particle Physics (CCEPP); Joint Large-Scale Scientific Facility Funds of the NSFC and CAS under Contracts Nos. U1532257, U1532258, U1732263; CAS Key Research Program of Frontier Sciences under Contracts Nos. QYZDJ-SSW-SLH003, QYZDJ-SSW-SLH040; 100 Talents Program of CAS; INPAC and Shanghai Key Laboratory for Particle Physics and Cosmology; German Research Foundation DFG under Contract No. Collaborative Research Center CRC 1044, FOR 2359; Istituto Nazionale di Fisica Nucleare, Italy; Koninklijke Nederlandse Akademie van Wetenschappen (KNAW) under Contract No. 530-4CDP03; Ministry of Development of Turkey under Contract No. DPT2006K-120470; National Science and Technology fund; The Swedish Research Council; U. S. Department of Energy under Contracts Nos. DE-FG02-05ER41374, DE-SC-0010118, DE-SC-0012069; University of Groningen (RuG); Helmholtzzentrum fuer Schwerionenforschung GmbH (GSI), Darmstadt; the Knut and Alice Wallenberg Foundation (Sweden) under Contract No. 2016.0157 and the Royal Society, UK under Contract No. DH160214.

%%%%%    bibliographies       Part                %%%%%%%%%%%%%
%%%%%%%%%%%%%%%%%%%%%%%%%%%%%%%%%%%%%%%%%%%%%%%%%%%%%%%%%%%%%%%%

\input{bibitem.tex}

\begin{figure}
  \centering
  \includegraphics[width=19cm]{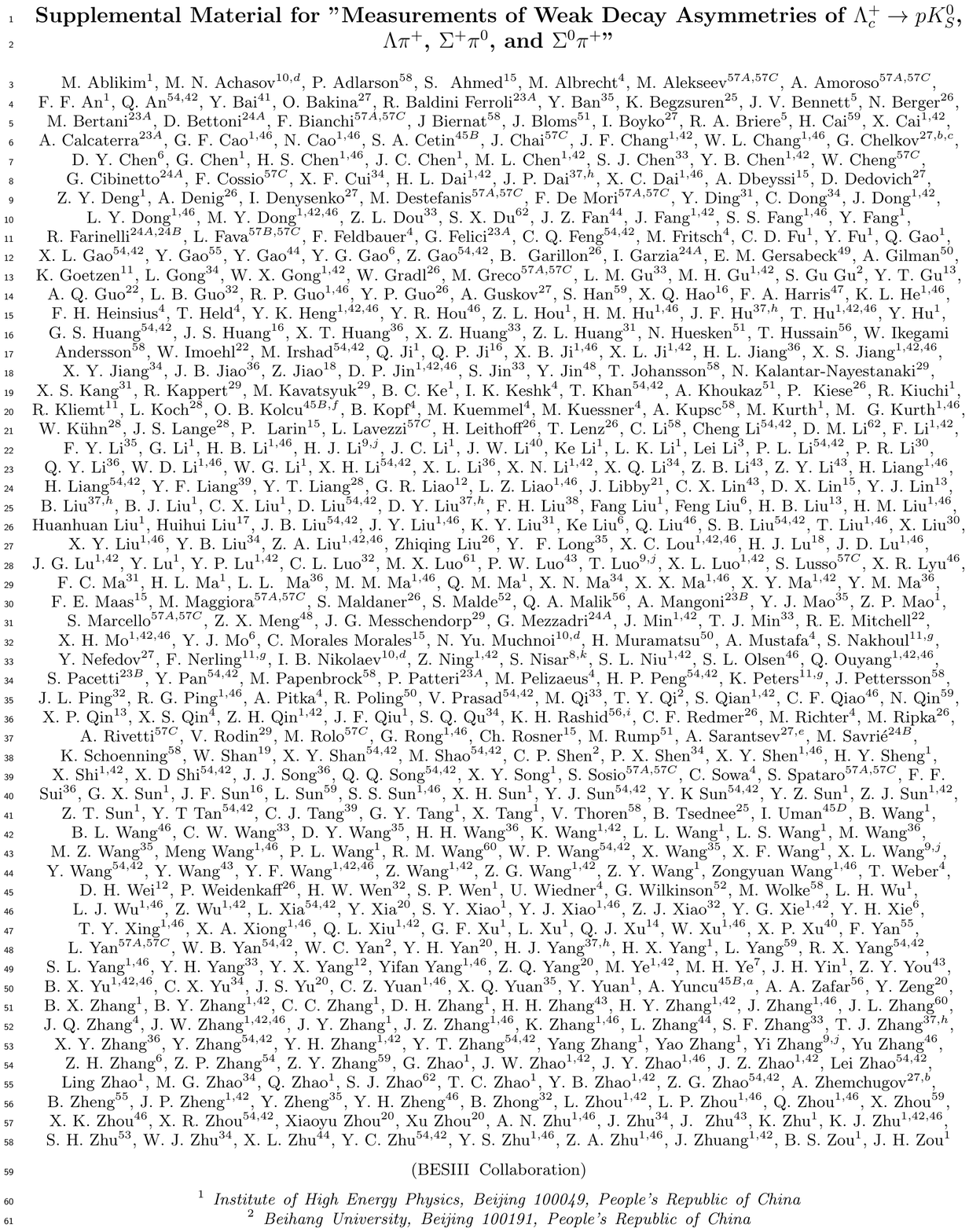}\\
\end{figure}
\begin{figure}
  \centering
  \includegraphics[width=19cm]{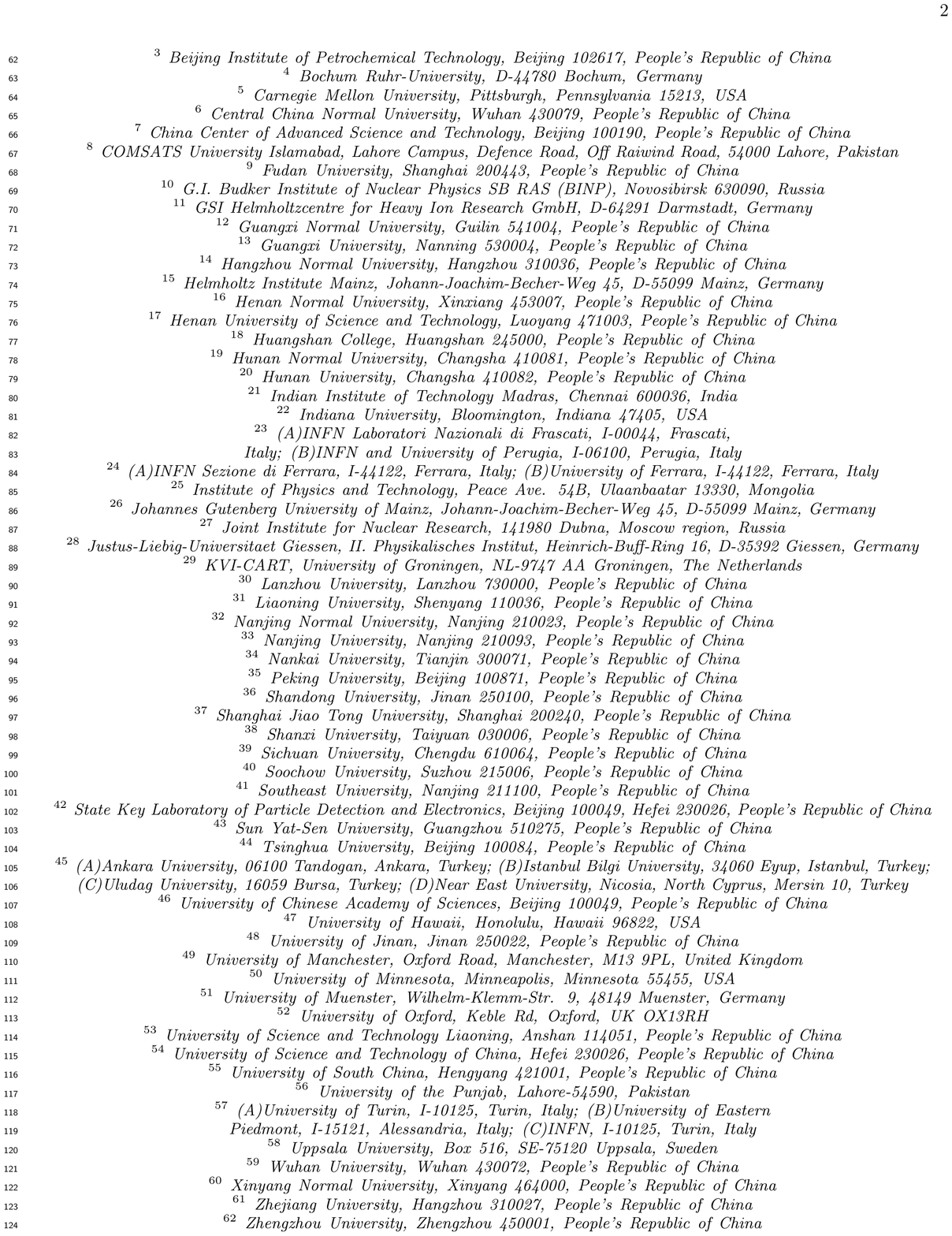}\\
\end{figure}
\begin{figure}
  \centering
  \includegraphics[width=19cm]{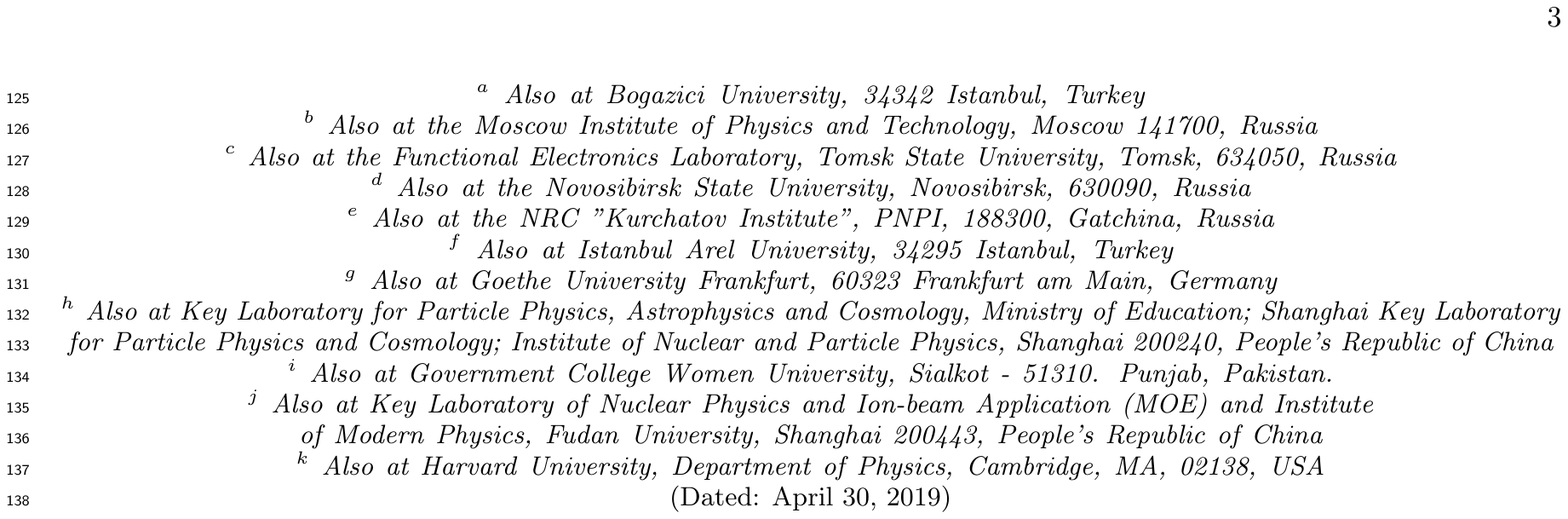}\\
\end{figure}
\begin{figure}
  \centering
  \includegraphics[width=19cm]{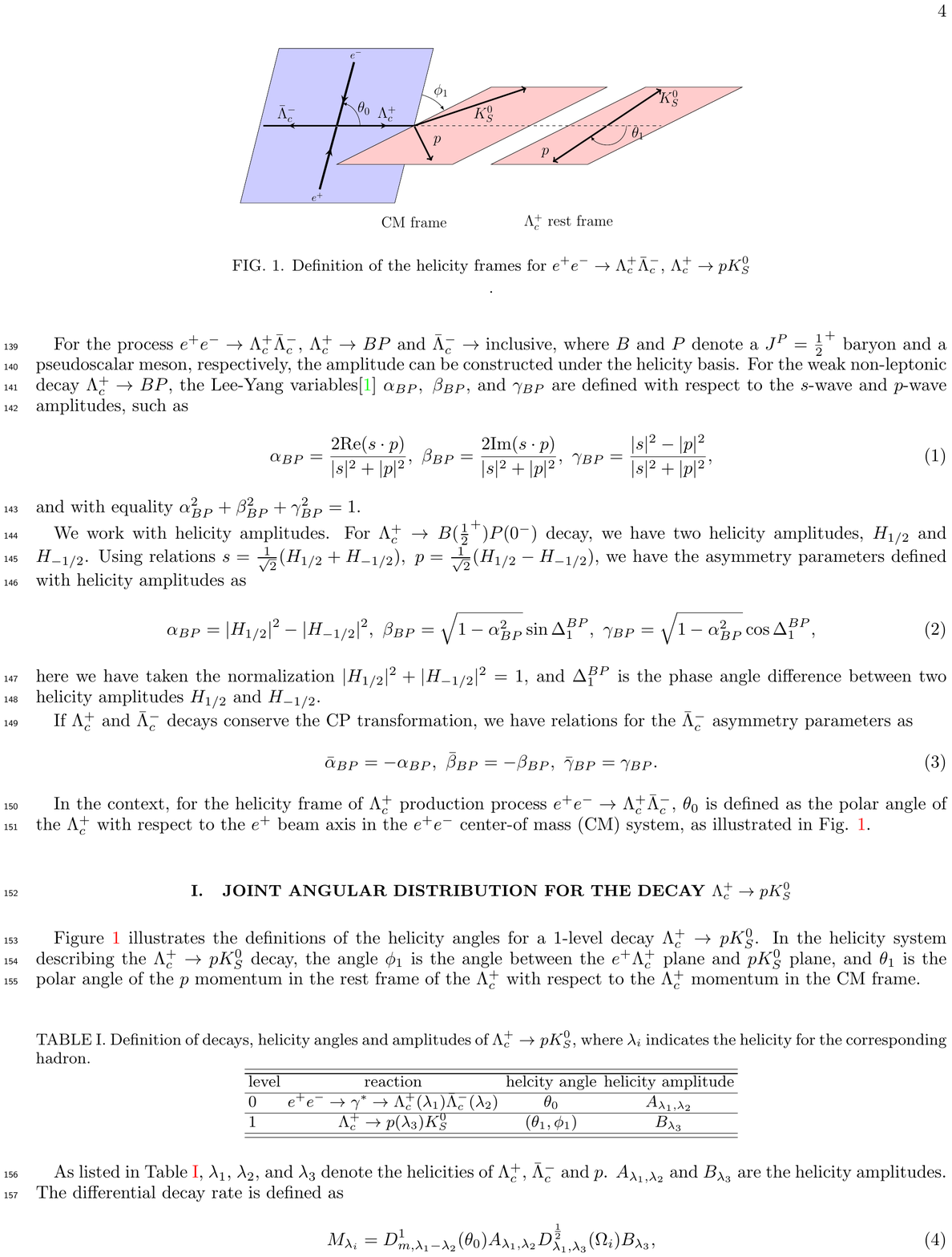}\\
\end{figure}
\begin{figure}
  \centering
  \includegraphics[width=19cm]{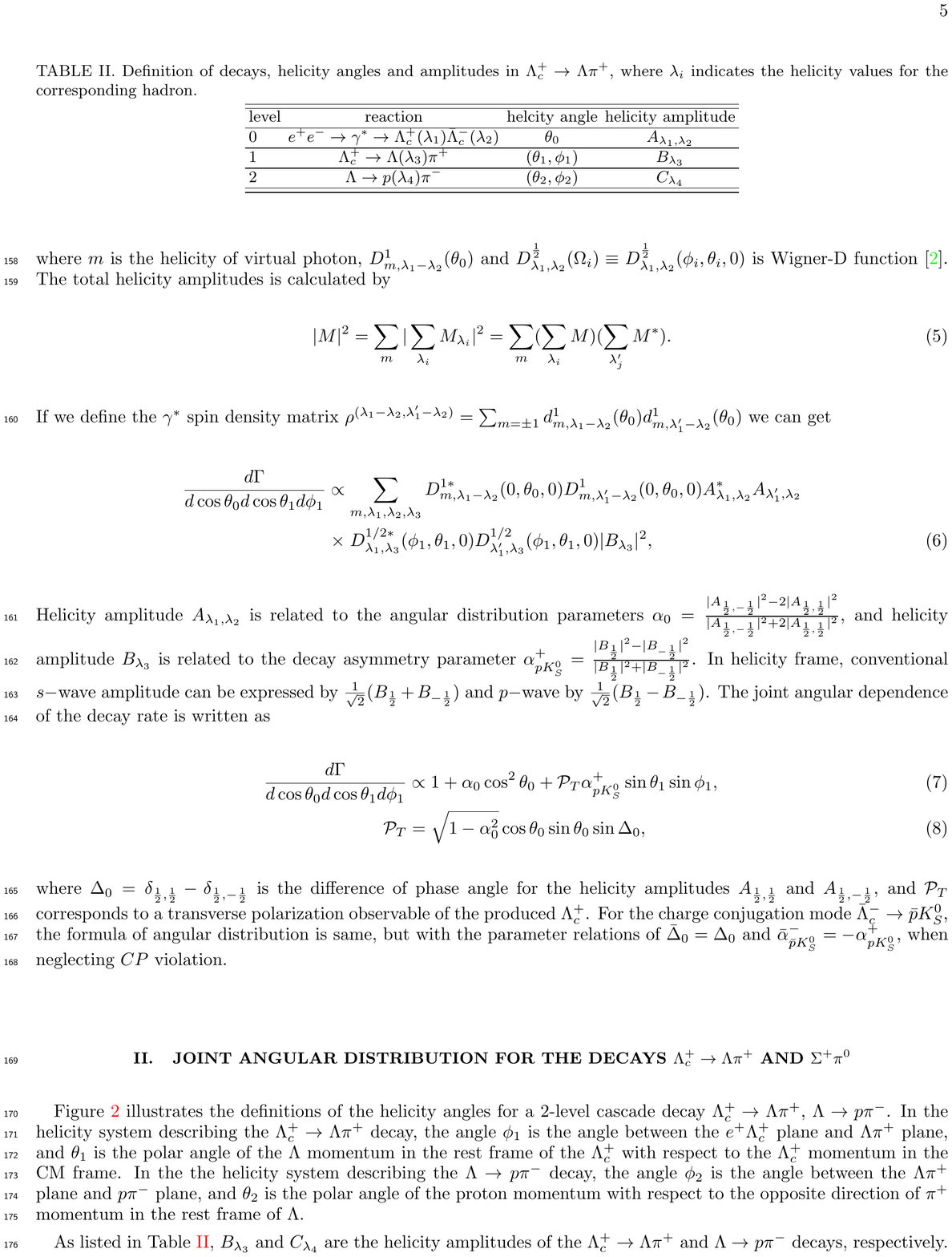}\\
\end{figure}
\begin{figure}
  \centering
  \includegraphics[width=19cm]{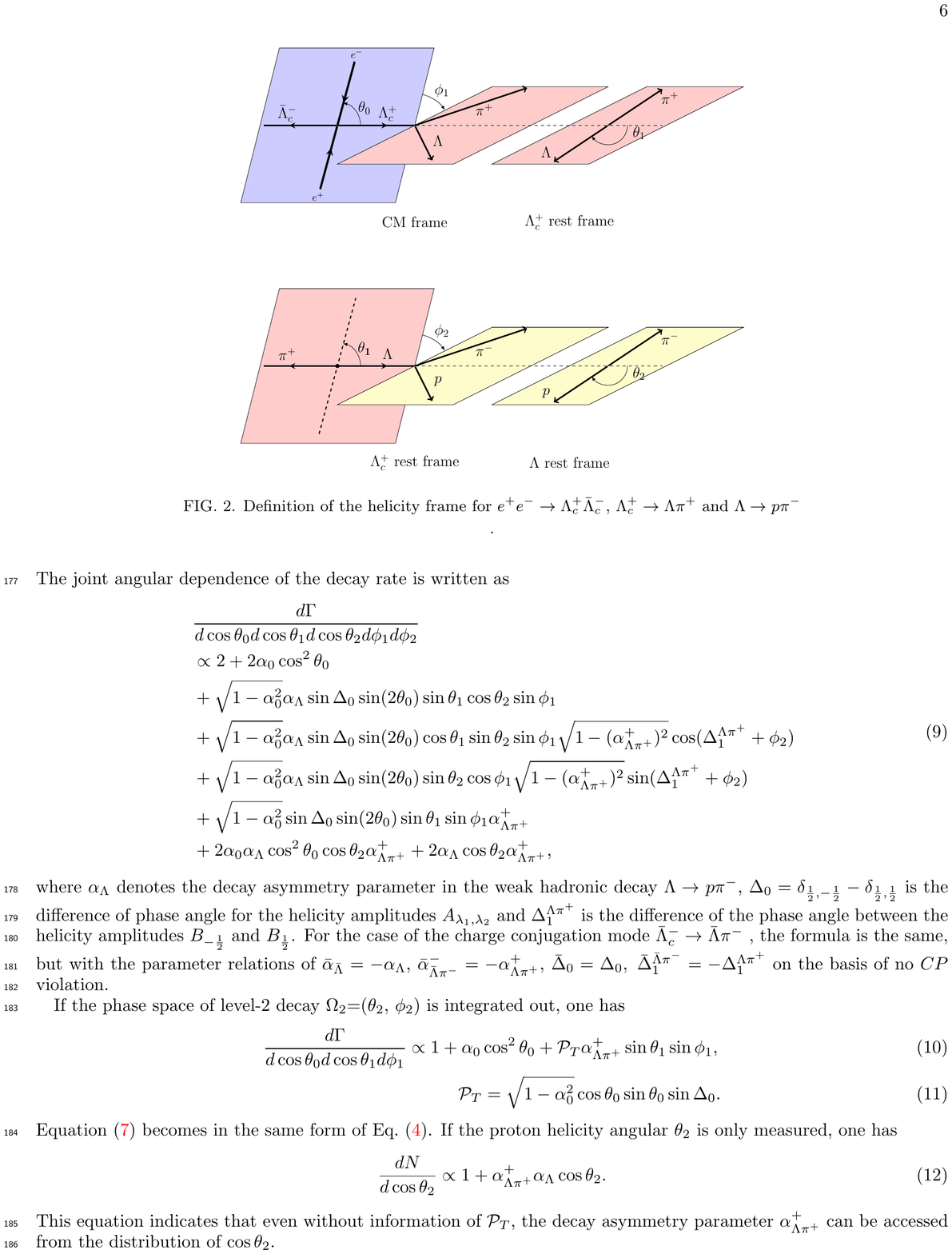}\\
\end{figure}
\begin{figure}
  \centering
  \includegraphics[width=19cm]{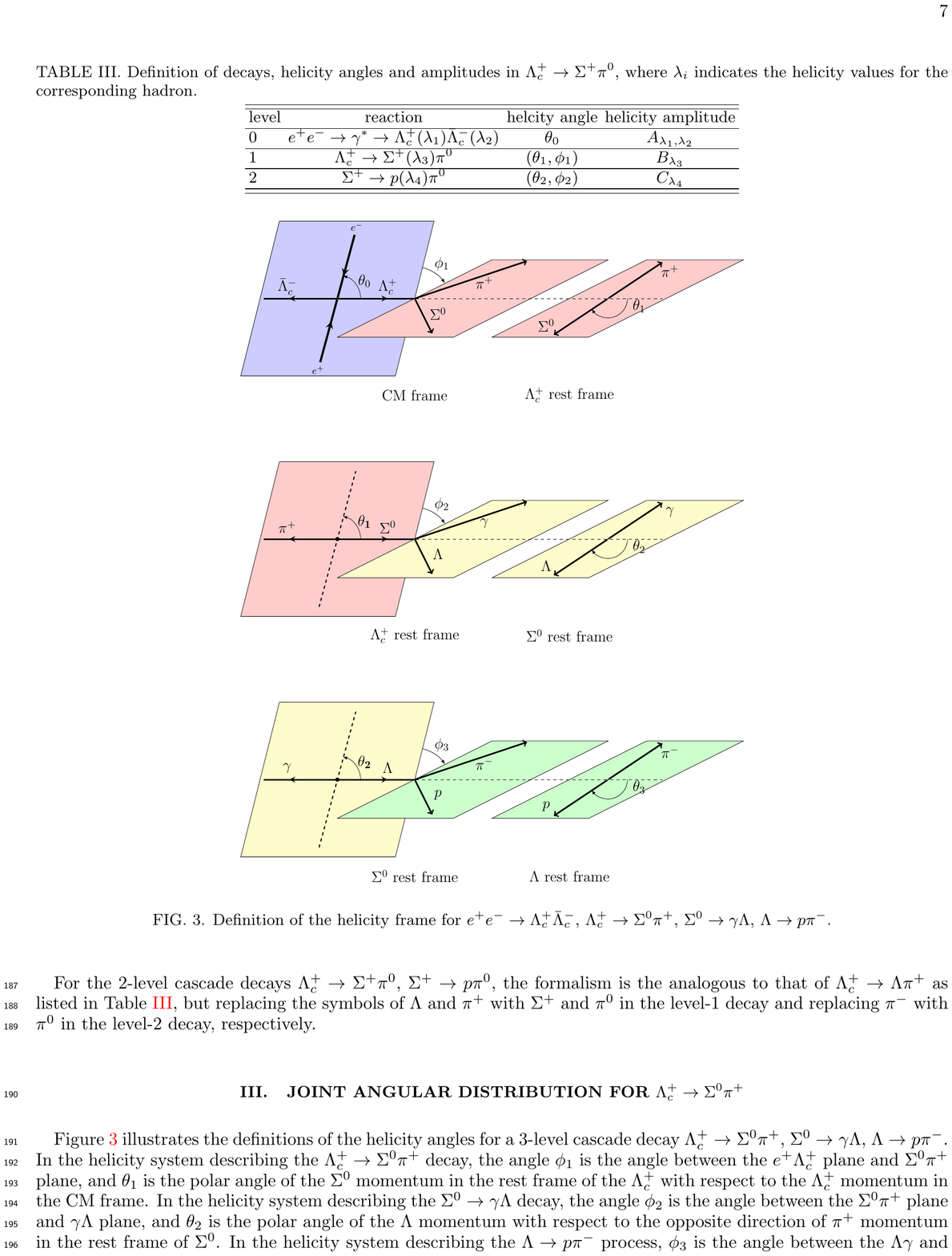}\\
\end{figure}
\begin{figure}
  \centering
  \includegraphics[width=19cm]{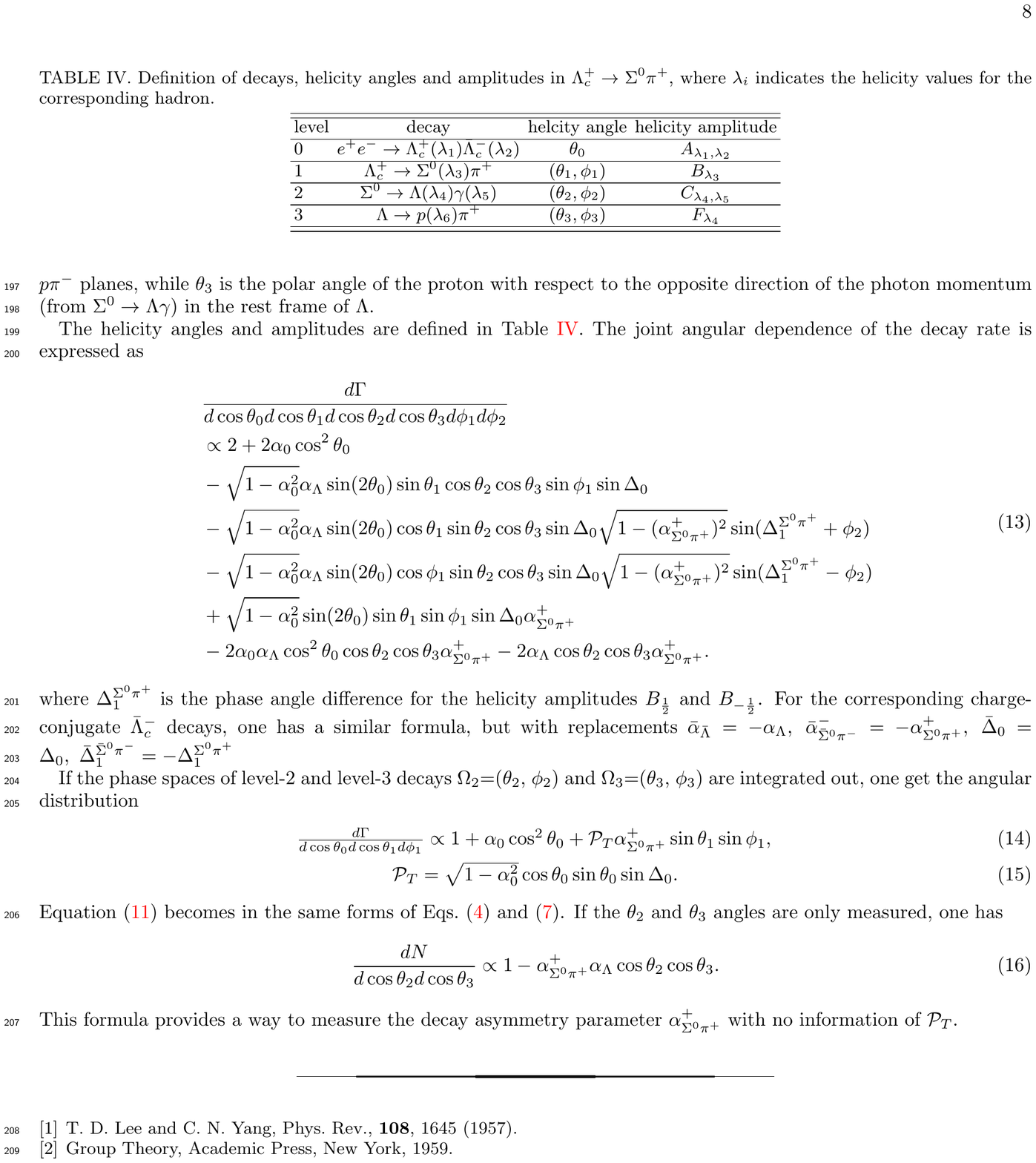}\\
\end{figure}

\end{document}

%% file: authors.tex
\author{
\begin{small}
M.~Ablikim$^{1}$, M.~N.~Achasov$^{10,d}$, P.~Adlarson$^{58}$, S. ~Ahmed$^{15}$, M.~Albrecht$^{4}$, M.~Alekseev$^{57A,57C}$, A.~Amoroso$^{57A,57C}$, F.~F.~An$^{1}$, Q.~An$^{54,42}$, Y.~Bai$^{41}$
% ...
\\
\vspace{0.2cm}
(BESIII Collaboration)\\
\vspace{0.2cm} {\it
$^{1}$ Institute of High Energy Physics, Beijing 100049, People's Republic of China\\
% ...
$^{4}$ Bochum Ruhr-University, D-44780 Bochum, Germany\\
% ...
$^{10}$ G.I. Budker Institute of Nuclear Physics SB RAS (BINP), Novosibirsk 630090, Russia\\
% ...
$^{15}$ Helmholtz Institute Mainz, Johann-Joachim-Becher-Weg 45, D-55099 Mainz, Germany\\
% ...
$^{41}$ Southeast University, Nanjing 211100, People's Republic of China\\
$^{42}$ State Key Laboratory of Particle Detection and Electronics, Beijing 100049, Hefei 230026, People's Republic of China\\
% ...
$^{54}$ University of Science and Technology of China, Hefei 230026, People's Republic of China\\
% ...
$^{57}$ (A)University of Turin, I-10125, Turin, Italy; (B)University of Eastern Piedmont, I-15121, Alessandria, Italy; (C)INFN, I-10125, Turin, Italy\\
$^{58}$ Uppsala University, Box 516, SE-75120 Uppsala, Sweden\\
% ...
$^{d}$ Also at the Novosibirsk State University, Novosibirsk, 630090, Russia\\
% ...
}
\vspace{0.4cm}
\end{small}
}

%%% Local Variables:
%%% mode: latex
%%% TeX-master: "omega-chicj"
%%% End:

%% file: bibitem.tex
%%%%%    bibliographies       Part                %%%%%%%%%%%%%
%%%%%%%%%%%%%%%%%%%%%%%%%%%%%%%%%%%%%%%%%%%%%%%%%%%%%%%%%%%%%%%%